\documentclass[10pt]{article}
\usepackage{authblk}
\usepackage[utf8]{inputenc}
\usepackage[english]{babel}
\usepackage{csquotes}
\usepackage[left=2.25cm,
            right=2.25cm,
            top=2.5cm,
            bottom=3cm
            ]{geometry}   
\usepackage{graphicx}
\graphicspath{ {/} }
\usepackage[utf8]{inputenc}
\usepackage{amssymb} 
\usepackage{physics}
\usepackage{amsthm}
\usepackage{amsmath}
\usepackage{float}
\usepackage{bbold}
\usepackage{kpfonts} 
\usepackage[T1]{fontenc}
\usepackage[backend=biber,style=numeric, sorting=none]{biblatex}
\usepackage[title]{appendix}
\newcommand{\lr}[1]{\left(#1\right)}

\newcommand{\pder}[2][]{\ensuremath{\frac{\partial#1}{\partial#2}}} 
\newcommand{\der}[2][]{\ensuremath{\frac{d#1}{d#2}}}

\newcommand{\avg}[1]{\mathbf{E}\left\{#1\right\}}
\newcommand{\bavg}[1]{\langle #1 \rangle}
\DeclareMathOperator{\sign}{sign}

\title{Statistical Mechanics of Learning via Reverberation in Bidirectional Associative Memories}
\author[1]{Martino Salomone Centonze}
\author[2]{Ido Kanter} 
\author[3]{Adriano Barra}

\affil[1,3]{Dipartimento di Matematica e Fisica, Università del Salento \& Istituto Nazionale di Fisica Nucleare, Lecce, Italy}
\affil[2]{Department of Physics, Bar-Ilan University, Ramat Gan, Israel }
\date{}

\begin{document}

\maketitle

\begin{abstract}
We study bi-directional associative neural networks that, exposed to noisy examples of an extensive number of random archetypes, learn the latter (with or without the presence of a teacher) when the supplied information is {\em enough}: in this setting, learning is heteroassociative --involving couples of patterns-- and it is achieved by reverberating the information depicted from  the examples through the layers of the network. By adapting Guerra's interpolation technique, we provide a full statistical mechanical picture of supervised and unsupervised learning processes (at the replica symmetric level of description) obtaining analytically phase diagrams, thresholds for learning, a picture of the ground-state in plain agreement with Monte Carlo simulations and signal-to-noise outcomes. In the large dataset limit, the Kosko storage prescription as well as its statistical mechanical picture provided by Kurchan, Peliti, and Saber in the eighties is fully recovered. Computational advantages in dealing with information reverberation, rather than storage, are discussed for natural test cases. In particular, we show how this network admits an integral representation in terms of two coupled restricted Boltzmann machines, whose hidden layers are entirely built of by grand-mother neurons, to prove that by coupling solely these grand-mother neurons we can correlate the patterns they are related to: it is thus possible to recover Pavlov's Classical Conditioning by adding just one synapse among the correct grand-mother neurons (hence saving an extensive number of these links for further information storage w.r.t. the classical autoassociative setting). 
\end{abstract}


\section{Introduction}
The fields of Artificial Intelligence (AI) and Statistical Mechanics (SM) have become increasingly closer since the seminal work of Amit, Gutfreund and Sompolinsky \cite{amit1985storing} who studied the Hopfield model of neural network (HNN) by inheriting {\em modi operandi} from the field of disordered statistical mechanics \cite{MPV}. 


Since then several properties of associative memories have been analysed, by studying several variations and extensions of the HNN. Examples include different support of spins and patterns (\emph{e.g.} real valued neurons instead of the usual boolean spin-like variables, in compact \cite{kuhn1993statistical} or unbounded support \cite{barra2018phase}), hierarchically correlated data \cite{krogh1988mean, bos1988martingale}, different neuronal structures such as bipartite and tripartite topologies \cite{smolensky1986information, fischer2012introduction, sejnowski1986higher,Agliari3RBM}, multi-node interactions (i.e., P-spin models) \cite{baldi1987number,Pedreschi1,Pedreschi2,krotov2016dense} or variation on the Hebbian prescription \cite{Dotsenko,dreaming,amitpnas,KanterAsy} just to mention a few of them (see also \cite{Lenka,Carleo} for recent reviews on the state of the art in  Machine Learning by a statistical mechanics perspective). 


However the Hopfield model, as many of its variations on theme, is {\em autoassociative}, meaning that providing the network a piece of a pattern, this triggers  retrieval of the entire pattern but solely of that pattern, while general intuition on memories is that they are {\em heteroassociative}, that is a fraction of a pattern can trigger retrieval of a bunch of patterns \cite{NatRev}. A typical example of heteroassociation, in the simplest setting involving solely two patterns, is  Pavlov's Classical Conditioning where perceiving a piece of a pattern drives the retrieval of a couple of patterns (as in the celebrated example starring a dog learning the correlation among the pattern {\em food} and the pattern  {\em ringing bell} \cite{Pavlov1}). A pure autoassociative framework a' la Hopfield can still account for Pavlov's Classical Conditioning \cite{PavlovNoi} (ultimately by relying upon the multi-tasking capabilities of the Hopfield model when provided with sparse datasets \cite{parallel1,parallel2,parallel3}), yet learning the correlation among the two patterns (that is, the bell and the food in the dog's example) requires updating an extensive number of synapses in that setting \cite{PavlovNoi}.
\newline
Indeed, soon after the Hopfield milestone \cite{HopPNAS}, Kosko created the Bidirectional Associative Memories (BAM) \cite{Kosko}, as a minimal heteroassociative generalization of the Hopfield model built of by two interacting visible layers that, relying upon reverberation rather than storage but preserving the Hebbian storage prescription, were able to retrieve couples of patterns, naturally mimicking Pavlov's phenomenon.  
\newline
To understand how Kosko reverberation takes place it is useful to remark that there is a behavioral equivalence among Hopfield neural networks (harmonic oscillator of biological learning) and Restricted Boltzmannm achines (building blocks of deep architectures in Machine Learning) \cite{barra2012equivalence,barra2018phase,Remi,Marc}: indeed it can be shown that (at least in the random setting) these two types of neural networks are subjected to the same thresholds for learning and give rise to the same pattern recognition capabilities\footnote{Strictkly speaking, the duality sharply holds for a RBM equipped with neurons in the hidden layer that are real-valued grand-mother cells equipped with a Gaussian prior  \cite{barra2012equivalence}.}  \cite{agliari2022emergence,supHNN}. Extending such a duality of representation to heteroassociations, we can easily understand how Kosko's learning via reverberation happens: the BAM network has indeed a natural dual representation in terms of two restricted Boltmann machines, whose visible layers are the layers of the original BAM, while their hidden layers are built of by grand-mother neurons\footnote{A grand-mother cell is an idealization of a neuron that is solely dedicated to recognition of a given pattern. Hence, if an RBM has to deal with $K$ patterns and it is equipped with a grand-mother cell setting, its hidden layer has to be built of by exactly $K$ hidden neurons and its output (in that layer) is a {\em hot-vector}, that is a vector with all its entries silenced but the one related to the pattern that the network is retrieving, e.g. a configuration of hidden neurons as $(-1,-1,-1,+1)$ implies that the RBM has to cope with four patterns and it is actually reconstructing the fourth pattern.} that are coupled as shown in Fig. \ref{Duality}. This duality allows to appreciate that connecting  two hidden grand-mother neurons (by a single synapses)  can save an extensive number of synapses w.r.t. standard (autoassociative) mechanisms in order to perform Pavlov-like retrieval, hence these networks acquire  a renewed interest nowadays, when much attention is paid to investigations about the role of synapses (as e.g. in the study of saliency maps toward eXplainable AI \cite{Fosca}).
\newline

While both the BAM \cite{BAM} and the RBM \cite{decelle2021restricted} are neural networks made of two-layers (i.e., these are bipartite spin glasses in a pure statistical mechanical jargon \cite{Bipartite}), they behave rather differently: RBM learning algorithm accomplishes gradient descent in the space of the weights \cite{HintonOld} (or variations on theme as e.g. Contrastive Divergence \cite{Contrastive}), while learning in the BAM is Hebbian, that is the former learns trough the differential path while the latter through the integral path\footnote{This duality within the statistical mechanical description of  neural networks share some conceptual similarities with the classical counterpart of analytical mechanics, where there are two main paths to reach the equations of motion (i.e. conveying essentially the same information), the former -the differential one- being the Lagrangian formulation of Mechanics, the latter -the integral one- being the Hamiltonian formulation of Mechanics.}. Further, in the BAM architecture both layers do not specialize in different tasks, they are meant to retrieve and, by retrieving one pattern per layer, they accomplish hetero-association phenomena.  RBM, instead, do not deal with couples of patterns and solely one of their layers is visible (and experiences the dataset), while the other is hidden and it is dedicated to inspect features and correlations hidden in the data exposed to the visible layer.
\newline
BAM's properties (at work with the simpler Hebbian storage prescription as in the original formulation, see also \cite{Peliti,Tirozzi}) have been recently analysed in detail  \cite{BAM} and plan for the present paper is to extend that analysis from heteroassociative Hebbian storage toward heteroassociative Hebbian learning (both supervised and unsupervised) as we briefly explain:  the synaptic structure of the BAM generalizes to couples of patterns the original proposal by Hebb ({\em neurons that fire together wire together}) \cite{hebb2005organization} that lies at the core of associative memories, providing a simple but powerful recipe that allows these models to retrieve the patterns of information. However, in the standard Amit-Gutfreund-Sompolinsky (AGS) theory \cite{amit1985storing}, the Hebb's rule is stated in terms of already defined patterns to be stored in the network's memory and typical question to answer is thus {\em which is the maximal storage capacity of these networks?}, that is Hebbian learning is a storage rule, not a genuine learning rule. However, driven by modern Machine Learning challenges, we would face an Hebbian learning where the network does not have direct access to these archetypes, while solely noisy versions of them (i.e. examples) are provided to it: beyond being more realistic w.r.t. the simpler Hebbian storing, this generalization allows to pose a number of new questions as, e.g., {\em which is the minimal amount of examples per archetype that the network has to experience, given the noise in these examples, the amount of archetypes to cope with and the available resources (neurons and synapses) for this task?}  or {\em do the synapses play all the same role or, for instance, synapses connecting grand-mother cells are intrinsically different w.r.t. synapses dealing with visible neurons?}
\newline
In this work, we try to answer these questions by proposing a Hebbian learning prescription  for the BAM (in both supervised and unsupervised settings) and performing an exhaustive study of these networks via disordered statistical mechanics.  The thermodynamics of the two models is studied by using the Guerra interpolation technique, at the replica symmetric level of approximation (RS), and corroborated by extensive Monte Carlo simulations. The phase diagrams are derived and both the learning as well as the retrieval performances of the BAMs are quantified in terms of standard order parameters (as the noise and the load of the network) but also in terms of novel control parameters, related to the dataset, as the ``dataset entropy'', that quantify the information provided to the network  (that is its {\em quality} and its {\em quantity}) and allows to derive explicitly the thresholds for learning: these new outputs of our statistical mechanics inspection return --at the leading orders--  power law critical scaling for the learning thresholds (as expected \cite{Kanter1,Kanter2,Kanter3}) and are also confirmed independently by a signal-to-noise investigation.
\newline
\newline
The paper is organized as follows: in Section \ref{SezioneDue} we introduce the settings, from the dataset and its information's content  \ref{Sezione2.1} to the network with its learning protocols \ref{Sezione2.2}. In Section \ref{SezioneTre} we provide the replica symmetric treatment of supervised learning in the BAM (in particular in Section \ref{Sezione3.1} Guerra's interpolation is introduced to solve the model, whose properties are analyzed in detail in the ground state \ref{Sezione3.2} and in the big-data limit \ref{Sezione3.3})  while Section \ref{SezioneQuattro} is dedicated to the unsupervised counterpart, mirroring the previous section. In Section \ref{SezioneCinque} we perform a signal-to-noise analysis to provide an independent, despite more raw, validation of the picture provided by statistical mechanics and finally Section \ref{SezioneSei} is left for conclusions and outlooks.

\section{The dataset, the neural network model and the training protocols}\label{SezioneDue}

Let us briefly introduce the Bidirectional Associative Memories (BAM) \cite{Kosko}.
\newline
Consider a shallow Hebbian network made of just two layers of neurons, the former built of by $N$ binary neurons $\sigma_i,\ i \in (1,...,N)$ the latter by $\overline{N}$  binary neurons $\tau_k,\ k \in (1,...,\overline{N})$ whose Cost function $H_N(\sigma,\tau|W)$ admits a Hamiltonian representation where the two layers of neurons mutually interact by the following prescription 
\begin{equation}
    H_N(\sigma,\tau|W) = - \sum_{i=1}^N \sum_{k=1}^{\overline N} w_{ik} \sigma_i \tau_k.
\end{equation}
The synaptic matrix $w_{ik}$ is symmetric (i.e. $w_{ik}=w_{ki}$) and aims to store couples of $N$-bit and  $\overline{N}$-bit long patterns $\xi^{\mu},\  \bar{\xi}^{\mu}$ (respectively)  solely within bidirectional interactions among neurons of different layers such that $W$ reduces to a linear combination of $K$ factorized terms, i.e. the \emph{patterns}, whose the explicit expression reads
\begin{equation}
    w_{ik} \propto \sum_{\mu=1}^K \xi^\mu_i \bar \xi^\mu_k,
    \label{eq:w}
\end{equation}
where the proportionality relation omits a normalization constant that ensures the Hamiltonian to be linearly extensive in the number of neurons. Eq. \ref{eq:w} constitutes the natural generalization of the Hebb's rule for storing single patterns to couples of patterns.
\newline
It is instructive to rephrase the above Cost function $H_N(\sigma,\tau|W)$ in terms of a standard Loss function of typical usage in Machine Learning:
\begin{equation}
    H_{N}(\sigma,\tau|W) = -\sqrt{N \overline N} \sum_{\mu} \lr{ 1 - \mathcal L_{N}^\mu}
\end{equation}
where the Loss function $\mathcal L_{N, \bar{N}}^\mu$ is defined in terms of the functions $L_{\sigma}^\mu$ and $\overline L_{\tau}^\mu$ as
\begin{align}
    &\mathcal L_{N, \bar{N}}^\mu = L_{\sigma}^{\mu} + \overline L_{\tau}^{\mu} - L^{\mu}_{\sigma} \overline L_{\tau}^{\mu},\label{eq:loss}\\
    &L_{\sigma}^{\mu} = \frac{1}{2N}\sum_i \lr{ \xi^{\mu} - \sigma_i }^2,\\
    &\overline L_{\tau}^{\mu} = \frac{1}{2\overline N}\sum_k \lr{ \overline \xi^{\mu} - \tau_k }^2.
\end{align}
We emphasize that, historically within the statistical mechanical formalization of neural networks, by writing  the Cost function $H_{N}(\sigma,\tau|W)$ we keep the focus on neural dynamics, e.g. to inspect the pattern recognition properties of the network, while when writing $\mathcal L_{N, \bar{N}}(W|\sigma,\tau)$ the focus is on the learning properties as emerging from synaptic dynamics. 

\subsection{The datasets and their conditional entropies}\label{Sezione2.1}
Extending the route paved in \cite{agliari2022emergence, supHNN} for the standard Hopfield model, in this paper we propose  a theory for bidirectional Hebbian learning, both in the supervised and unsupervised scenarios (namely in presence -or absence- of a teacher): roughly speaking this means that the network is no longer exposed directly to the patterns (that it never experiences), rather, solely noisy examples of these patterns are provided to the network that has thus to infer correctly the archetypes hidden in these corrupted versions once {\em enough information} has been collected.
\newline
In this context, the terms $\xi^\mu, \bar \xi^\mu$, as defined in Eq. \ref{eq:w}, are promoted to play as \emph{archetypes} and used to generate a dataset $\mathcal D = \{\eta^a, \bar \eta^a \:|\: a=1,..,M\}$ of related \emph{examples}: the index $a=1,..,M$ runs over the number of examples available for each archetype, that is assumed to be the same for all of them.

We define the \emph{examples} $\eta$ and $\overline \eta$ as the noisy version of the \emph{archetypes} $\xi$ and $\overline \xi$ by randomly flipping fractions of the original pixels via 
\begin{align}
    &\eta^{\mu a}_i = \xi^\mu_i \chi^{\mu a}_i, \label{eq:eta}\\
    &\overline \eta^{\mu b}_k = \bar \xi^\mu_k \bar \chi^{\mu b}_k. \label{eq:etabar}
\end{align}
Following the standard assumptions of AGS theory \cite{amit1985storing}, the \emph{archetypes} $\xi\:(\bar \xi)$ are random vectors  whose entries assume values $\{\pm 1\}^{N}$ ($\{\pm 1\}^{\overline N}$ respectively) with probability $1/2$. The $\chi, \overline \chi$ flipping-noises distribute according to 
\begin{align}
    &\mathcal P (\chi_a) =  p \delta (\chi_a -1) + (1-p) \delta (\chi_a +1),\\
    &\mathcal P (\bar \chi_a) = \bar p \delta (\bar \chi_a -1) + (1-\bar p) \delta (\bar \chi_a +1).
\end{align}
such that the first momentum of the distributions of $\chi, \bar \chi$ reads
\begin{align}
    &\avg{\chi}_\chi = 2p -1 \equiv r, \:\:
    \avg{\bar \chi}_{\bar \chi} = 2\bar p -1 \equiv \bar r.
\end{align}
Hence, in the limit of $r,\bar r \to 1$ (i.e. $p,\bar p \to 1$), the examples $\eta, \overline \eta$ collapse to the archetypes $\xi, \overline \xi$ (and consequently, the entropy of the related dataset $\mathcal D$ vanishes), while, in the other limit  $r,\bar r \to 0$ (i.e., $p,\bar p \to 1/2$), the examples become random vectors totally uncorrelated to their archetypes (and consequently, the entropy of the related dataset is therefore maximal). 
\newline
Notice that, at difference with standard (autoassociative) Hebbian learning,  in the present context, mixed situations -where the noise level of the dataset is not the same in the examples $\eta$ and $\overline \eta$- can arise\footnote{This is the typical situation in datasets collected from fluorescence microscopy where the emission of the fluorescence molecules tackling the signals have different intensities \cite{PNASnoi,pixel}.}.
\newline
The following statistical properties of the \emph{examples} $\eta, \bar \eta$ can be consequently traced:
\begin{align}
    &\avg{\frac{1}{M} \sum_a \eta^{\mu a}_i}_{\eta} = 0,\\
    &\avg{\frac{1}{M} \sum_a \bar\eta^{\mu a}_i}_{\bar\eta} = 0,\\
    &\avg{\left(\frac{1}{M} \sum_a \eta^{\mu a}_i\right)^2}_{\eta} = r^2 + \frac{1-r^2}{M} \equiv R,\\
    &\avg{\left(\frac{1}{M} \sum_a \bar\eta^{\mu a}_k\right)^2}_{\bar\eta} = \bar r^2 + \frac{1-\bar r^2}{M} \equiv \bar R,\\
    &\avg{\left(\frac{1}{M} \sum_a \eta^{\mu a}_i \bar\eta^{\mu a}_k \right)}_{\eta, \bar\eta} = r^2 \bar r^2 + \frac{1- r^2 \bar r^2}{M} = r^2 \bar r^2 \lr{1+\rho}.
\end{align}
In the last equation, we have introduced the parameter
\begin{align}
    \rho= \frac{1-r^2\bar r^2}{Mr^2\bar r^2},\label{eq:rho_unsup}
\end{align}
which will be central in the analysis of the unsupervised model.\\
The information content of the original message carried by the archetypes that is retained by the examples is quantified for each pattern $\mu$ and digit $i,k$, $(\xi^\mu_i$,$\bar \xi^\mu_k)$, and block $(\eta_i^\mu,\bar\eta_k^\mu) = \left(\sum_a \eta_i^{\mu a},\sum_a \bar\eta_k^{\mu a}\right)$, through the conditional entropy $S(\xi^\mu_i,\bar\xi^\mu_i|\eta^\mu_i, \bar\eta^\mu_i)$:
\begin{align}
    S(\xi^\mu_i,\bar\xi^\mu_i|\eta^\mu_i, \bar\eta^\mu_i) = \mathbf E_{\sim P(\xi^\mu_i,\bar\xi^\mu_i, \eta^\mu_i,\bar\eta^\mu_i)} \log P(\xi^\mu_i,\bar\xi^\mu_i | \eta^\mu_i,\bar\eta^\mu_i). 
\end{align}
In the following, we avoid repeating the indices $i,k,\mu$ in the equations in order to keep the notation minimal.
Notice that the pair of examples $(\eta,\bar\eta)$ are statistically independent, as they are only related to the independent archetypes $\xi$ and $\bar \xi$ (i.e., $P(\xi,\bar\xi)=P(\xi)P(\bar\xi)$ thus $P(\eta,\bar\eta)=P(\eta)P(\bar\eta)$). This consideration allows us to express the joint probability as $P(\xi,\bar\xi, \eta,\bar\eta) = P(\xi,\eta) P(\bar\xi,\bar\eta)$. From this follows that the conditional entropy defined above is the sum of two independent terms, one per layer, that is
\begin{align}
    S(\xi,\bar\xi|\eta, \bar\eta) = S( \xi|\eta) + S(\bar \xi|\bar \eta).\label{eq:condentropy}
\end{align}
Consider the computation of $S(\xi|\eta)$. The examples $\eta$ are given by eq. (\ref{eq:eta}) and are therefore determined by the quantity $\sum_a \chi^a$. One can distinguish between two cases, depending on the sign of the product $\xi \eta = sgn(\sum_a \chi^a) = \pm 1$. Hence, the conditional entropy $S( \xi|\eta)$ can be written as
\begin{align}
    S( \xi|\eta) = - P_+ \log P_+ - P_-\log P_-,
\end{align}
where 
\begin{align}
    &P_+ = \mathcal P(sgn(\sum_a \chi^a)= 1) = \mathcal P(\sum_a \chi^a \geq 0) \sim \frac{1}{2}(1+\erf (1/\sqrt{2\rho_s}), \:\:\: M \to \infty,\\
    &P_- = \mathcal P(sgn(\sum_a \chi^a)= -1) = \mathcal P(\sum_a \chi^a < 0) \sim \frac{1}{2}(1-\erf (1/\sqrt{2\rho_s}), \:\:\: M \to \infty.
\end{align}
The last step is justified by virtue of the Central Limit Theorem (CLT). 
\newline
Note that the conditional entropy only depends on the parameter $\rho_s$, defined as:
\begin{align}
    \rho_s = \frac{1-r^2}{Mr^2},
\end{align}
and it is an increasing function of $\rho_s$. The computation of $S(\bar \xi|\bar \eta)$ is identical, and naturally leads to the definition of $\bar \rho_s$:
\begin{align}
    \bar \rho_s = \frac{1-\bar r^2}{M\bar r^2}.
\end{align}
Hence, the parameters $\rho_s,\bar \rho_s$ uniquely determine the conditional entropy (\ref{eq:condentropy}) of the dataset: at $(\rho_s,\bar \rho_s) = (0,0)$, the conditional entropy of the dataset is zero, therefore the examples coincide with the archetypes; at $(\rho_s,\bar \rho_s) \to (\infty,\infty)$ the conditional entropy is maximal and the examples are no longer representative of the original archetypes.
\newline
As a consequence, w.r.t. the standard AGS theory on Hebbian storing, here we have more control parameters, namely $M, \bar{M}$ and $r, \bar{r}$ (that are the control parameters related to the datasets, quantifying its {\em quantity} and {\em quality}), or we can consider as effective control parameters directly the entropies of the datasets that they ruled, that is $\rho_s$ and $\bar{\rho_s}$. The remaining (standard) control parameters, related to the network's load and noise, are introduced in the next section.

\subsection{The neural networks and their learning protocols}\label{Sezione2.2}

The direct consequence of dealing with examples rather than patterns is that the Hebbian storage rule for reverberation \cite{BAM} has to be extended toward a generalized Hebbian learning rule for reverberation and this gives rise to two scenarios: the unsupervised and the supervised settings. 
\newline
In the case of unsupervised Hebbian learning, the examples are directly summed up in the Hamiltonian of the model, resulting in the following form of the related synaptic matrix:
\begin{equation}
    w_{ik}^{unsup} \propto \sum_{\mu=1}^K \sum_{a=1}^M \eta^{\mu a}_i \bar \eta^{\mu a}_k.
    \label{eq:unsupw}
\end{equation}
This expression implements the idea that a teacher -able to distinguish \emph{a priori} the different categories (i.e. the different archetypes in our jargon) the examples belong to- is not available, hence the best we can do is to supply directly to the network the whole available information  as a linear superposition of all of the examples.\\
In the supervised counterpart instead,  due to the help of a teacher, we do distinguish among different categories (i.e. different archetypes) before supplying the collected information to the network:  the natural choice for the synaptic matrix is thus
\begin{equation}
    w_{ik}^{sup} \propto \sum_{\mu=1}^K  \sum_{a=1}^M \eta^{\mu a}_i \sum_{b=1}^M \bar \eta^{\mu b}_k.
    \label{eq:supw}
\end{equation}
In the following we investigate the information processing capabilities of the BAM in both cases, tackling the problem by a statistical mechanics perspective.
In order to do so, hereafter we share the basic definitions that are in order: first, we explicitly introduce the two Cost functions that contextualize the BAM within an unsupervised (supervised) protocol. 
\newline
\newline
The Cost function of the BAM network trained without the presence of a teacher -that is, under an unsupervised protocol- is given by 
\begin{equation}
    H_{N,M}^{unsup}(\sigma,\tau|\eta)=-\frac{1}{ML\sqrt{R\overline R}} \sum_{\mu, a = 1}^{K,M} \sum_{i,k = 1}^{N,\overline N} \eta^{\mu a}_i \bar\eta^{\mu a}_k \sigma_{i} \tau_k ,
    \label{eq:unsup}
\end{equation}
while the Cost function of the BAM network trained under the help of a teacher -that is, within a supervised protocol- reads as 
\begin{equation}
    H_{N,M}^{sup}(\sigma,\tau|\eta)=-\frac{1}{LM^2 \sqrt{R\overline R}} \sum_{\mu=1}^K \sum_{i,k=1}^{N,\overline N} \lr{\sum_{a=1}^M \eta^{\mu a}_i} \lr{ \sum_{b=1}^M \bar \eta^{\mu b}_k} \sigma_i\tau_k,
    \label{eq:sup}
\end{equation}
where we have chosen the geometric mean $L=\sqrt{N\overline N}$ as the normalization coefficient (instead of the total number of neurons $N+\overline N$), preserving notation consistency w.r.t. \cite{BAM}. 
\newline
Note that, while standard artificial learning in e.g. restricted Boltzmann machines \cite{decelle2021restricted} (i.e. trained within the {\em contrastive divergence} class of algorithms \cite{Contrastive}), feed-forward networks \cite{HopNN} (i.e. trained via the {\em back propagation} class of chain rules \cite{BackProp}), etc. typically involves evaluation of the gradients of the Loss functions w.r.t. the weights (as discussed in the incipit of this Section), hence it works trough evaluation of derivatives, at contrary 
Hebbian learning -that is an oversimplified representation of biological learning- takes place by the integral path, that is by taking explicit summations over the examples (as directly coded in the related Cost functions) so to take advantage of concentration of measure arguments as those provided by the classical convergence dictat of CLT.

Back to the introduction of the standard statistical mechanical package of definitions, once provided the Hamiltonians (i.e the Cost functions) and introduced the fast noise (i.e, the {\em temperature} T) in the network ruled by the control parameter $\beta  \equiv T^{-1} \in \mathcal{R}^+$,  the associated partition function $Z_{N,\bar{N},M,\bar{M}}(\beta)$ reads
\begin{equation}
    Z_{N,\bar{N},M,\bar{M}}(\beta) = \sum_{\{\sigma\} \{\tau\}}^{2^N, 2^{\overline{N}}} e^{-\beta H},
\end{equation}
and allows to define the main observable, namely the quenched free energy $-\beta f(\beta)= A(\beta)$, where the quenching is performed by the average $\mathbb{E}_{\eta,\overline \eta}$ over the variables $\eta,\overline \eta$ after taking the logarithm of the partition function:
\begin{equation}\label{eq:Adefinition}
    A(\beta)=\lim_{N, \bar{N} \to \infty} \frac{1}{L}\mathbb{E}_{\eta,\overline \eta}\ln  Z_{N,\bar{N},M}(\beta).   
\end{equation}
We aim to obtain an explicit expression of the quenched free energy in terms of the natural order parameters of the theory: these are the archetype magnetizations $m^\mu_\sigma$ and $m^\mu_\tau$ and the two-replica overlaps $q^{\alpha \beta}$ and $\bar q^{\alpha \beta}$ (one per layer):
\begin{align}
    &m^\mu_\sigma = \frac{1}{N} \sum_{i=1}^N \xi^\mu_i \sigma_i,\\
    &m^\mu_\tau = \frac{1}{\overline N} \sum_{k=1}^{\overline N} \bar \xi^\mu_k \tau_k,\\
    &q^{\alpha \beta} = \frac{1}{N} \sum_{i=1}^N  \sigma_i^\alpha\:\sigma_i^\beta,\\
    &\bar q^{\alpha \beta} = \frac{1}{\overline N} \sum_{k=1}^{\overline N} \tau_k^\alpha\:\tau_k^\beta,
\end{align}
where the indices $\alpha,\beta$ are the replica indices. Further, in the unsupervised setting, we have also the magnetizations relative to the examples $\eta^a,\bar \eta^a$, indicated as $n^{\mu a}_\sigma$ and $n^{\mu a}_\tau$, whose definitions are 
\begin{align}
    &n^{\mu a}_\sigma = \frac{1}{r\sqrt{(1+\rho_s)(1+\bar \rho_s)}}\frac{1}{N} \sum_{i} \eta^{\mu a}_i \sigma_i,\label{eq:exmagnetizations_unsupsigma}\\
    &n^{\mu a}_\tau = \frac{1}{\bar r\sqrt{(1+\rho_s)(1+\bar \rho_s)}}\frac{1}{\overline N} \sum_{k,a} \bar \eta^{\mu}_k \tau_k,\label{eq:exmagnetizations_unsuptau}    
\end{align}
while, in the supervised counterpart, the relevant example's magnetization order parameters are relative to the average examples $\sum_a \eta^{\mu a}$ and $\sum_a \bar \eta^{\mu a}$ and defined as
\begin{align}
    &n^\mu_\sigma = \frac{1}{r\sqrt{(1+\rho_s)(1+\bar \rho_s)}}\frac{1}{NM} \sum_{i,a} \eta^{\mu a}_i \sigma_i,\label{eq:exmagnetizations_supsigma}\\
    &n^\mu_\tau = \frac{1}{\bar r\sqrt{(1+\rho_s)(1+\bar \rho_s)}}\frac{1}{\overline N M} \sum_{k,a} \bar \eta^{\mu a}_k \tau_k.\label{eq:exmagnetizations_suptau}\\
\end{align}
\newline
Beyond $\beta$ -that accounts for the noise level within the network- the control parameters related to the network's load and architecture $\alpha$, $\bar \alpha$ and $\gamma$ are instead defined as
\begin{align}
    &\alpha = \lim_{N\to\infty} K/N,\\
    &\bar \alpha = \lim_{\overline N \to\infty}K/\overline N = \alpha \gamma^2,\\
    &\gamma = \lim_{N,\overline N \to\infty} \sqrt{N/\overline N},
\end{align}
where $\alpha$ and $\bar{\alpha}$ account for the network's load per layer  while $\gamma$ tunes the asymmetry in the network's layer sizes.
We define also the `global' load of the network $\lambda$ as
\begin{align}
    \lambda = \lim_{L\to\infty} K/L = \sqrt{\alpha \bar \alpha} = \alpha \gamma.
\end{align}
The plan is to achieve an explicit expression of the quenched free energy in terms of the control and order parameters, so to extremize the former w.r.t. the latter: this standard procedure results in a series of (self-consistent) equations that rule the evolution of the order parameters in the space of the control parameters, whose study allows to draw phase diagrams, namely plots in the space of the control parameters where different regions depict different emergent computational capabilities collectively shown by these assemblies of neurons and synapses.
\newline
\newline
A first remark here is that, while in the classical domain of usage of statistical mechanics (i.e., when the Hamiltonian represents a physical system) this plan guarantees thermodynamics to hold, in this extension of statistical mechanics to neural network, it can be seen as the extremization of the cost function under the maximum entropy inferential perspective, as developed by Jaynes \cite{Jaynes}: in these regards the cost function acts as a Lagrangian multiplier to the maximum entropy inferential criterion and the whole free energy can be seen as a  complex constrained optimization problem \cite{Ksat1,Ksat2}.
\newline
\newline
A second remark consists in observing that phase diagrams, the standard final output of the statistical mechanical journey in the neural network's world, may play nowadays a pivotal role {\em en route} toward Sustainable and Optimized AI (SAI $\&$ OAI) \cite{KaoCO2} because the a-priori knowledge of the regions in the control parameters where the network loses some information processing capabilities allows to save energy and CPU time badly assigned to mismatched problems and, at contrary, forcing the network in the regions of the phase diagram where the desired properties can be achieved allows a conscious usage of neural networks\footnote{One could still argue that the bulk of analytical calculations required to obtain phase diagrams is performed on random data-sets, while real data-sets contain structure. While this is certainly true, yet the Shannon compression argument guarantees that if there is a not-empty region where a given computational capability is spontaneously shown by the network, such a region would eventually enlarge (due to potential compression of structured information), but not diminish, hence the random setting is universal and works as a lower-bound picture for structured scenarios, that have to be addressed one-by-one.}.
\newline
\newline
In the following two Sections we investigate supervised and unsupervised learning protocols (one per Section) achieved by reverberation in the BAM networks. 
\newline
As standard in the bulk of statistical mechanical works of neural networks \cite{Coolen}, we develop the theory at the {\em replica symmetric} (RS) level of description: roughly speaking we assume that all the order parameters concentrate on their means in the infinite volume limit, that is
$$
\lim_{N,\bar{N} \to \infty} P(x) = \delta(x-\langle x \rangle)
$$
where $x$ is a generic order parameter, function of the $N$ neurons $\sigma$ and/or $\bar{N}$ neurons $\tau$, $P(x)$ is its p.d.f. and $\langle x \rangle$ represents its average.

\section{Replica Symmetric Theory of Supervised Learning via Reverberation}\label{SezioneTre}

As the couples of archetypes share all the same importance (a straight consequence of the RS scenario we are working in), we focus (with no loss of generality) solely on the first couple (and their related examples): this represent the signal, while all the remaining information  constitutes the (slow) noise playing against its learning, storage and retrieval.

To handle the post-synaptic potential it is convenient to rely upon the mean field nature of these networks: this allows to apply generalized Carmona-Wu universality theorems regarding the quenched noise in spin glasses \cite{Carmona,Genovese} to treat these slow noise contributions as Gaussian mixtures: by introducing the quenched  Gaussian variables $\lambda, \bar \lambda$, as proved in \cite{Pedreschi1,Pedreschi2}, by virtue of  CLT we can write 
\begin{align}
    &\sum_a \eta^{\mu a}_i \sim M R^{1/2} \lambda^\mu_i, \:\: \lambda \sim \mathcal N (0,1)\\
    &\sum_a \bar \eta^{\mu b}_k \sim M \bar R^{1/2} \bar \lambda^\mu_k, \:\: \bar\lambda \sim \mathcal N (0,1),
\end{align}
and it is convenient to re-define the statistical pressure (i.e. the quenched free energy) as
\begin{align}
    A(\beta) = \lim_{L\to\infty} \frac{1}{L} \avg{\ln Z(\beta)},
\end{align}
where the expectation is now performed on the effective variables $\lambda,\ \bar \lambda$ and those accounting for the signal, that are $\eta^1,\ \bar \eta^1$.

\bigskip

Keeping in mind the distinction between the signal ($\mu=1$) and the (slow) noise ($\mu>1$) and taking advantage of the integral representation of a Gaussian\footnote{As the Cost function we are considering are pairwise, the {\em energy} they represent are quadratic forms such that the exponential of the Cost function depicts a Gaussian shape.} we write the partition function as
\begin{align}\label{IntegraZ}
    Z_{N,\bar{N},M}(\beta) =\sum_{\{\sigma\} \{\tau\}}^{2^N,2^{\bar{N}}} \exp\left(\frac{\beta }{LM^2\sqrt{R \overline R}} \sum_{i,k=1}^{N,\overline N} \eta^{1}_i \bar \eta^{1}_k \sigma_i \tau_k \right)\int Dz^\dag Dz \exp\left( \frac{\beta}{ \sqrt{N}} \sum_{i,\mu>1} \lambda_i^\mu z_\mu \sigma_i + \frac{\beta}{ \sqrt{\overline N}} \sum_{k,\mu>1} \bar \lambda_k^\mu z^\dag_\mu \tau_k \right),
\end{align}
where $\eta^{1}_i,\bar \eta^{1}_k$ are short-hands for $\eta^{1}_i = \lr{\sum_{a=1}^M \eta^{1 a}_i}$ and $\bar \eta^{1}_k=\lr{\sum_{a=1}^M \bar \eta^{1 a}_k}$. The difference w.r.t. the standard Hopfield case \cite{barra2012equivalence} is that now the variables $z, z^\dag$ are \emph{i.i.d.} coupled complex variables distributed according to the integral measure given by
\begin{align}\label{Pavlov}
    Dz^\dag Dz = \prod_{\mu>1} dz_\mu^\dag dz_\mu \exp\left( -\beta z_\mu^\dag z_\mu \right).
\end{align}
We can therefore introduce two further order parameters, the conjugated momenta $p$ and $\overline p$, defined as follows:
\begin{align}
    &p^{\alpha\beta} = \frac{r}{\bar r}\sqrt\frac{1+\rho_s}{1+\bar \rho_s} \frac{1}{K-1} \sum_{\mu>1} z_\mu^\alpha\:z_\mu^\beta,\label{eq:sup_p}\\
    &\bar p^{\alpha\beta} = \frac{\bar r}{r}\sqrt\frac{1+\bar \rho_s}{1+ \rho_s}  \frac{1}{K-1} \sum_{\mu>1} (z^\dag_\mu)^\alpha\:(z^\dag_\mu)^\beta.\label{eq:sup_pbar}\\
\end{align}
Equation (\ref{IntegraZ}) follows from the identity:
\begin{equation}
    \prod_{\mu>1} \exp \left( \beta' n^\mu_\sigma n^\mu_\tau \right) = \left( \frac{\beta'}{\pi} \right)^{K-1} \int Dz^\dag Dz\: \exp\left(\beta' \sum_{\mu>1} n_\sigma^\mu z_\mu + \beta' \sum_{\mu>1} n_\tau^\mu z^\dag_\mu\right).
\end{equation}
\begin{figure}[!ht]    
    \centering
    \includegraphics[width=16.5cm]{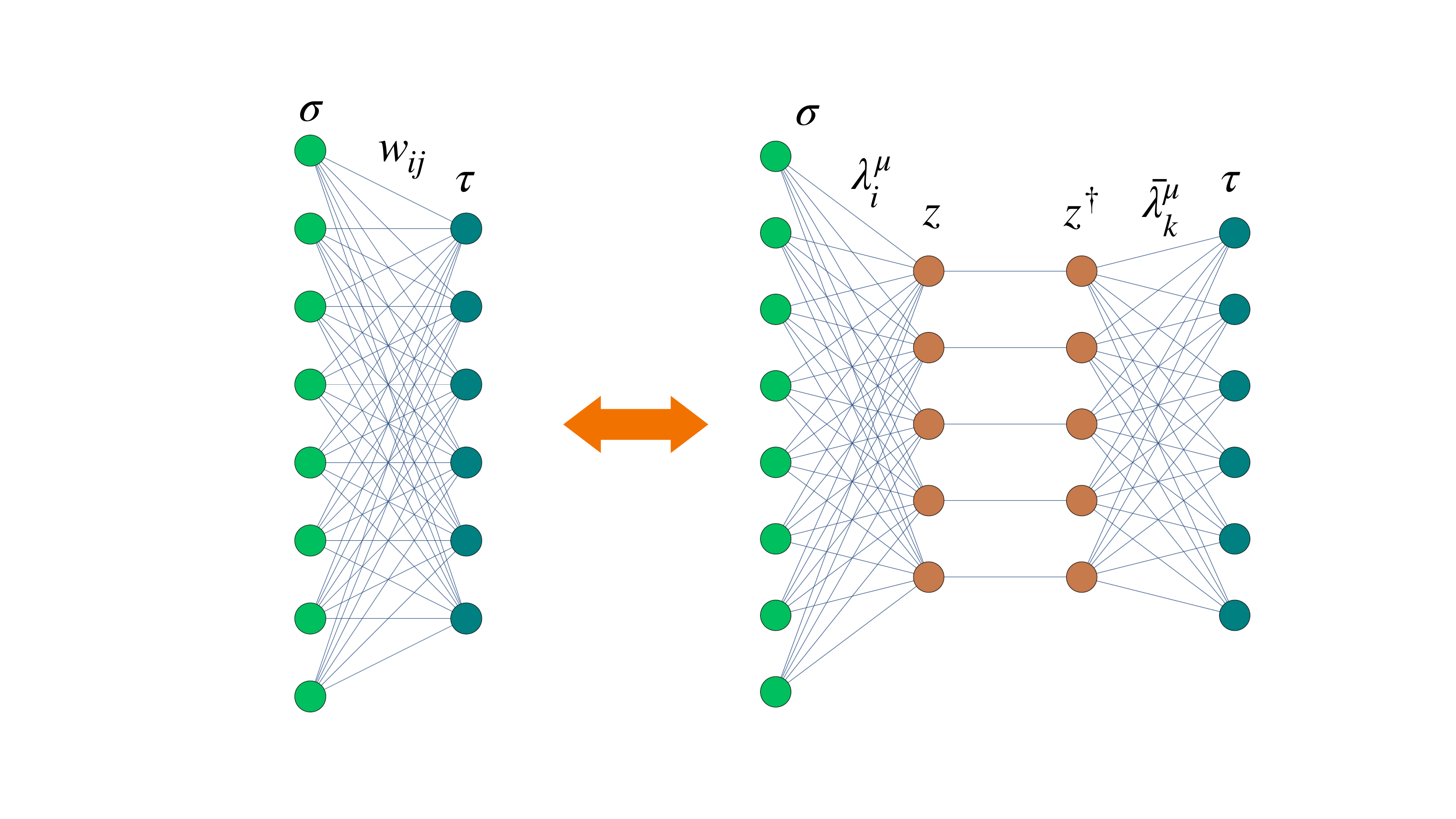}
    \caption{\label{Duality} Dual representation of the Bidirectional Associative Memories. In the left plot, the representation in terms of just visible neurons is shown: note that, in this setting, the weights are Hebbian. In the right plot, the integral representation in terms of two Restricted Boltzmann machines, whose hidden neurons are grand-mother cells mutually coupled among the two hidden layers. The coupling among hidden grand-mother neurons is expressed by eq. (\ref{Pavlov}): the latter guarantees, in the direct model provided in the left picture, that the retrieval of one concept on a visible layer is paired to the retrieval of its conjugate concept on the opposite visible layer.}
\end{figure}
An important remark toward the development of an optimized theory for neural networks is in order here: the integral representation of the partition function provided in eq. (\ref{IntegraZ}) shows that the BAM network can be seen as two interacting restricted Boltzmann machines, where the former has the visible layer $\{\sigma \}$ interacting with the hidden layer $\{z \}$ while the latter has the visible layer $\{\tau \}$ interacting with the hidden layer $\{z^\dag \}$. Further, these RBMs operate within the so-called {\em grand-mother cell setting} \cite{agliari2022emergence,grandmother} namely there is one and just one hidden neuron $z_{\mu}$ in the first RBM and one and just one hidden neuron $z_{\mu}^{\dag}$ in the second RBM such that, each time an example pertaining to archetype $\mu$ is provided to the visible layer $\{ \sigma \}$, the corresponding hidden neuron $z_{\mu}$ fires and, likewise, each time an example pertaining to archetype $\mu^\dag$ is provided to the visible layer $\{ \tau \}$, the corresponding hidden neuron $z_{\mu}^\dag$ fires as well: crucially, these two RBMs are not independent, rather they are coupled by the term coded in eq. (\ref{Pavlov}) and it is this coupling among $z_{\mu}$ and $z_{\mu}^\dag$ to allow the retrieval of couples of patterns, instead of single patterns as in standard Hebbian settings (e.g. the Hopfield model).
\newline
It is thus remarkable that, in order to correlate two patterns, while performing Classical Conditioning solely on visible neurons implies the update of an extensive number of synapses (the longer the patterns to relate, the larger the amount of required synapses to handle their correlation) \cite{PavlovNoi}, by connecting directly  the grand-mother cells (that is the hidden neurons $z$ and $z^\dag$) accounting for these patterns, we are linking neurons highly selective  to specific archetypes (and not those that code for their pixels) hence we save an extensive amount of synapses w.r.t. the auto-associative counterpart,  yet preserving the same computational scenario:  this highlights the heterogeneous role played by synapses in these artificial neural networks.

\subsection{Supervised Guerra's interpolation}\label{Sezione3.1}


We are now ready to tackle the explicit calculation of the quenched free energy. To this task we aim to apply Guerra's interpolation technique (see e.g. \cite{broken} for its genesis in spin glasses and \cite{GuerraHop} for its applications to neural networks): the standard plan when using Guerra's approach is to introduce an interpolating parameter $t\in(0,1)$ and an interpolating free energy whose extrema lie between the original model (already split in a signal and a noise term, as they appear in eq. (\ref{IntegraZ})) -that we achieve by setting $t=1$-  and a linear combination of factorized one-body terms, that we code in the auxiliary function $\phi(t)$, whose analytical treatment is feasible and that we achieve by setting $t=0$. We then solve for the latter, namely we evaluate the free energy at $t=0$, and then propagate back the solution at $t=1$ via the Fundamental Theorem of Calculus. The whole is reported in the following scheme:
\begin{align}\label{InterMilan}
    &A(\beta)=A(\beta; t=1) = A(\beta; t=0) + \int_0^1 dt' \frac{d}{dt'} A(\beta; t'),\\
    &A(\beta; t) = \lim_{L\to\infty} \frac{1}{L} \avg{\ln Z(\beta; t)},\\
    &\frac{d}{dt} A(\beta; t) = \lim_{L\to\infty} \frac{1}{L} \langle \frac{d}{dt} Z(\beta; t) \rangle_t,
\end{align}
where the brackets represent $\langle . \rangle = \mathbb{E}\{\omega(.)\}$, namely the quenched $\mathbb{E}$ expectation of the Boltzmann average $\omega(.)$.\\
We use to define a generalized interpolating partition function that reads as
\begin{align}
    &Z_{N,\bar{N},M}(\beta;t)=\sum_{\{\sigma\} \{\tau\}}\int Dz^\dag Dz \:e^{-\beta W(t)}.
\end{align}
Notice that the $e^{-\beta W}$ measure automatically defines the modified Boltzmann and quenched averages as, considering a generic function of the neurons $f(\sigma, \tau)$, we can write 
\begin{align}
 &   \omega_t(f(\sigma, \tau)) \equiv \sum_{\{\sigma\} \{\tau\}}\int Dz^\dag Dz \:f(\sigma, \tau) \: e^{-\beta W(t)},\\ 
  & \langle f(\sigma, \tau)\rangle_t \equiv \avg {\omega_t(f(\sigma, \tau))},
\end{align}
where $\omega_t(.)$ denotes the generalized Boltzmann average, which depends on $t$ and reduces to the standard one of statistical mechanics whenever evaluated at $t=1$. Likewise,  $ \langle f(\sigma, \tau) \rangle := \avg {\omega(f(\sigma, \tau))}$.\\
The details of the computation are shown in Appendix \ref{supAppendix}. The expression for the quenched \emph{RS} statistical pressure $A_{RS}$ reads
\begin{align}
    A_{RS}= &-\beta \sqrt{R\overline R}\: n_\sigma n_\tau - \frac{\beta^2\alpha \gamma }{2} \left(  p (1-q) +  \bar p (1-\bar q)\right)+\nonumber\\
    &-\frac{\alpha \gamma}{2} \ln \left( 1-\beta^2 (1-q)(1-\bar q) \right) + \frac{\alpha \gamma \beta^2}{2} \frac{ q(1-\bar q) + \bar q (1-q)}{1-\beta^2 (1-q)(1-\bar q)}+\nonumber\\
    &+\gamma \mathbf{E}_{\chi}\int Dx\: \ln \cosh \left( \beta \sqrt{\alpha   p} \:x + \beta \bar r n_\tau \gamma^{-1} \left( \frac{1}{M} \sum_a \chi_a\right) \right)+\nonumber\\
    &+\gamma^{-1} \mathbf{E}_{\bar \chi}\int Dx\: \ln \cosh \left( \beta \sqrt{\bar \alpha  \bar p}\: x + \beta r n_\sigma \gamma \left( \frac{1}{M} \sum_a \bar \chi_a\right) \right),
    \label{eq:ARS2}
\end{align}
At given values of the control parameters ($\alpha, \beta, \gamma, \rho_s, \bar{\rho_s}$), the \emph{RS} quenched pressure is a function of the order parameters, namely the magnetizations $n_\sigma, n_\tau$ and  the overlaps $q, \bar q, p, \bar p$ and we look for its extrema, w.r.t. these order parameters, that are thus obtained by requiring that its gradient vanishes. 
\newline
This request results in the following the self-consistencies equations for the order parameters:
\begin{align}
    &n_\sigma = \mathbf{E}_{\chi} \left( \frac{1}{M} \sum_a \chi_a\right) \int Dx\: \tanh \lr{\beta \sqrt{\alpha  p} \:x + \beta \frac{n_\tau \gamma^{-1}}{r} \left( \frac{1}{M} \sum_a \chi_a\right)},\label{eq:nsigma}\\
    &n_\tau = \mathbf{E}_{\bar \chi} \left( \frac{1}{M} \sum_a \bar \chi_a\right) \int Dx\: \tanh \lr{\beta \sqrt{\bar \alpha \bar p}\: x + \beta \frac{n_\sigma \gamma}{\bar r} \left( \frac{1}{M} \sum_a \bar \chi_a\right)},\label{eq:ntau}\\
    &q = \mathbf{E}_{\chi} \int Dx\: \tanh^2 \lr{\beta \sqrt{\alpha  p} \:x + \beta \frac{n_\tau \gamma^{-1}}{r}\left( \frac{1}{M} \sum_a \chi_a\right)},\label{eq:q}\\
    &\bar q = \mathbf{E}_{\bar \chi} \int Dx\: \tanh^2 \lr{\beta \sqrt{\bar \alpha \bar p}\: x + \beta \frac{n_\sigma \gamma}{\bar r} \left( \frac{1}{M} \sum_a \bar \chi_a\right)},\label{eq:barq}\\
    &p = \frac{\bar q + \beta^2 q (1-\bar q)^2}{\left( 1-\beta^2 (1-q)(1-\bar q) \right)^2},\label{eq:p}\\
    &\bar p = \frac{q + \beta^2 \bar q (1-q)^2}{\left( 1-\beta^2 (1-q)(1-\bar q) \right)^2}.\label{eq:pbar}
\end{align}
Further, the expected archetype magnetizations $\bavg{m^1_\sigma},\: \bavg{m^1_\tau}$, can be formally derived by relying upon the functional generator trick, that is by adding two currents of the form $N J m_\sigma + \overline N \bar J m_\tau$ to the functional $W$ defined within the partition function $Z$. This results in a new term in the argument of both $\ln\cosh$ functions in the $A_{RS}$ functional in Eq. (\ref{eq:ARS1}), namely $J \xi$ and $\bar J \bar \xi$ in the third and fourth line respectively. Then, one simply has to evaluate
\begin{align}
    \langle m_\sigma \rangle = \frac{1}{N}\der[A_{RS}]{J}\biggr\rvert_{J=0}, \:\:\:\:\:\: 
    \langle m_\tau \rangle = \frac{1}{\overline N}\der[A_{RS}]{\bar J}\biggr\rvert_{\bar J=0}.
\end{align}
The self-consistency equations for the archetype magnetizations read as 
\begin{align}
    &m_\sigma = \mathbf{E}_{\chi} \int Dx\: \tanh \lr{\beta \sqrt{\alpha  p} \:x + \beta \frac{n_\tau \gamma^{-1}}{r} \left( \frac{1}{M} \sum_a \chi_a\right)},\label{eq:msigma}\\
    &m_\tau = \mathbf{E}_{\bar \chi} \int Dx\: \tanh \lr{\beta \sqrt{\bar \alpha \bar p}\: x + \beta \frac{n_\sigma \gamma}{\bar r} \left( \frac{1}{M} \sum_a \bar \chi_a\right)}\label{eq:mtau}.
\end{align}

\subsection{Big-Data Regime ($M \to \infty$ limit)}\label{Sezione3.2}
One may argue that the construction to infer the archetype magnetization is somehow artificial, but, in general (that is for arbitrarily small dataset sizes $M$) there are no other routes. At contrary, in the {\em big-data limit}  $M\to \infty$ (namely when the number of examples diverges as well with the size of the system $N, \bar{N}$), the mean example magnetization and the archetype magnetization become directly related as we now prove. 
\newline
At first we can rewrite the following quantities by virtue of the CLT, introducing the stochastic variables $s,\bar s$, as:
\begin{align}
    &\frac{1}{M} \sum_a \chi_a \to r + \sqrt{\frac{1-r^2}{M}} \: s = r(1+\sqrt \rho_s\: s),  \:\:  \:\:  \:\: s \in \mathcal N (0,1), \:\:  \:\:  \:\: M\to \infty \:\: ,\label{eq:fluctuations_1}\\
    &\frac{1}{M} \sum_a \bar\chi_a \to r + \sqrt{\frac{1-\bar r^2}{M}} \: \bar s = \bar r(1+\sqrt {\bar \rho_s} \: \bar s),  \:\:  \:\:  \:\: \bar s \in \mathcal N (0,1), \:\:  \:\:  \:\: M\to \infty \:\: \label{eq:fluctuations_2},
\end{align}
where we used the control parameters $\rho_s, \bar \rho_s$.
The self-consistency equations (\ref{eq:nsigma}-\ref{eq:barq}), (\ref{eq:msigma}-\ref{eq:mtau}) can be re-written in terms of these control parameters, and, using the following property of Gaussian integrals, for any $a,b,c \in \mathbb{R}$:
\begin{align}
    \int Dx Dy \:f(a x + b y +c) = \int Dz\: f\left(\sqrt{a^2+b^2}\:z + c\right),
\end{align}
we can drop the dependence from $s,\bar s$ in the equations. The self-consistency equations then read:
\begin{align}
    &n_\sigma = \frac{1}{\sqrt{(1+\rho_s)(1+\bar \rho_s)}} \frac{m_\sigma + \beta (1-q) m_\tau \gamma^{-1}\frac{\rho_s}{\sqrt{(1+\rho_s)(1+\bar \rho_s)}}}{1-\beta^2(1-q)(1-\bar q)\frac{\rho_s\bar \rho_s}{(1+\rho_s)(1+\bar \rho_s)}},\label{eq:ns_sup}\\
    &n_\tau = \frac{1}{\sqrt{(1+\rho_s)(1+\bar \rho_s)}} \frac{m_\tau + \beta (1-\bar q) m_\sigma \gamma\frac{\bar \rho_s}{\sqrt{(1+\rho_s)(1+\bar \rho_s)}}}{1-\beta^2(1-q)(1-\bar q)\frac{\rho_s\bar \rho_s}{(1+\rho_s)(1+\bar \rho_s)}},\label{eq:nt_sup}\\
    &m_\sigma = \int Dx\: \tanh \left( \beta \sqrt{\alpha p + (n_\tau \gamma^{-1})^2 \rho_s}\:x + \beta n_\tau \gamma^{-1} \right),\label{eq:ms_sup}\\
    &m_\tau = \int Dx\: \tanh \left( \beta \sqrt{\bar \alpha \bar p + (n_\sigma \gamma)^2 \bar \rho_s}\:x + \beta n_\sigma \gamma\right),\label{eq:mt_sup}\\
    &q = \int Dx\: \tanh^2 \left( \beta \sqrt{\alpha p + (n_\tau \gamma^{-1})^2 \rho_s}\:x + \beta n_\tau \gamma^{-1} \right),\label{eq:q_sup}\\
    &\bar q = \int Dx\: \tanh^2 \left( \beta \sqrt{\bar \alpha \bar p + (n_\sigma \gamma)^2 \bar \rho_s}\:x + \beta n_\sigma \gamma \right),\label{eq:qbar_sup}\\
    &p = \frac{\bar q + \beta^2 q (1-\bar q)^2}{\left( 1-\beta^2 (1-q)(1-\bar q) \right)^2},\label{eq:p_sup}\\
    &\bar p = \frac{q + \beta^2 \bar q (1-q)^2}{\left( 1-\beta^2 (1-q)(1-\bar q) \right)^2}\label{eq:pbar_sup}.
\end{align}

\begin{figure}[!ht]    
    \centering
    \includegraphics[width=7.5cm]{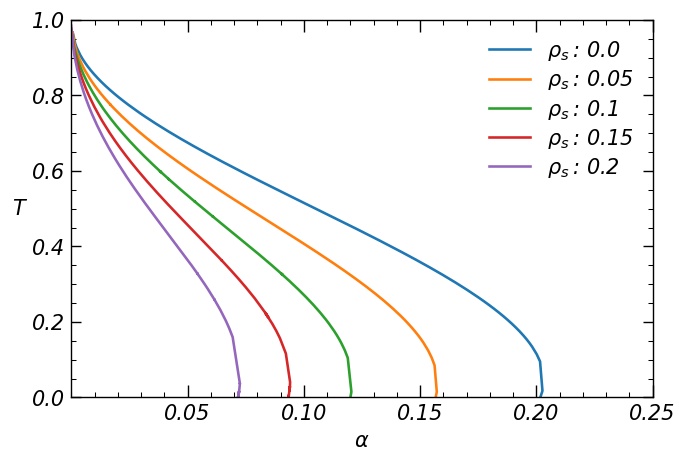}
    \includegraphics[width=7.5cm]{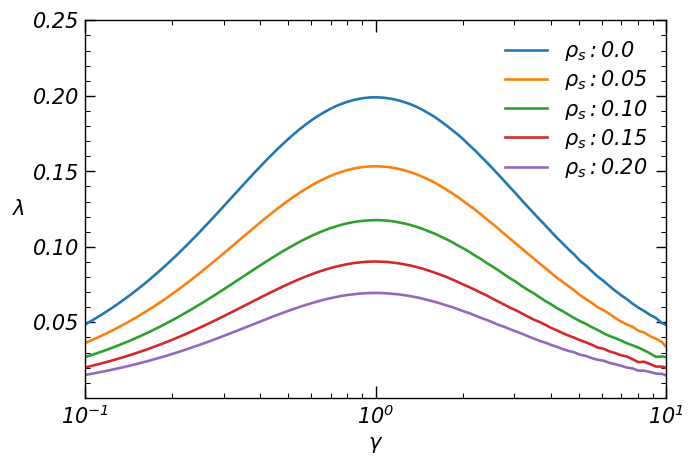}
    \caption{Left: the critical lines dividing the retrieval phase of the model from the spin glass phase are displayed for various values of $\rho_s= \bar \rho_s$ and $\gamma = 1$. Right: the critical global load $\lambda = \alpha \gamma$ is shown for different values of $\rho_s= \bar \rho_s$ as a function of $\gamma$.}
    \label{Fig:PD_sup}
\end{figure}

\begin{figure}[!ht]    
    \centering
    \includegraphics[width=5.6cm]{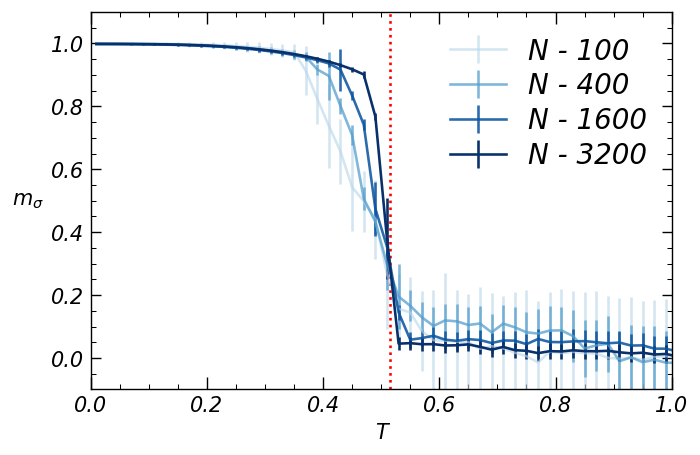}
    \includegraphics[width=5.6cm]{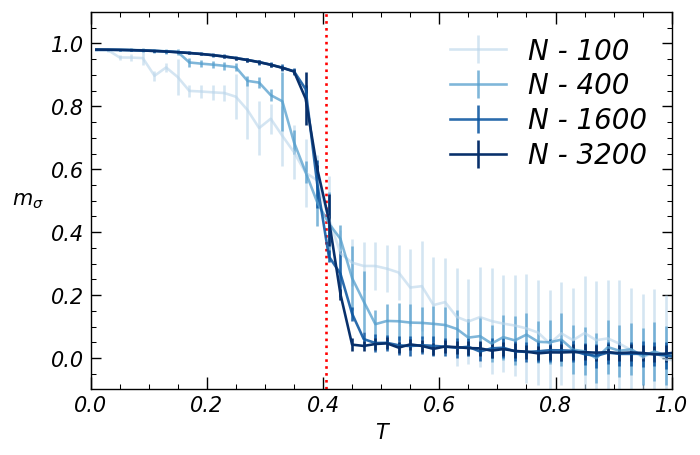}
    \includegraphics[width=5.6cm]{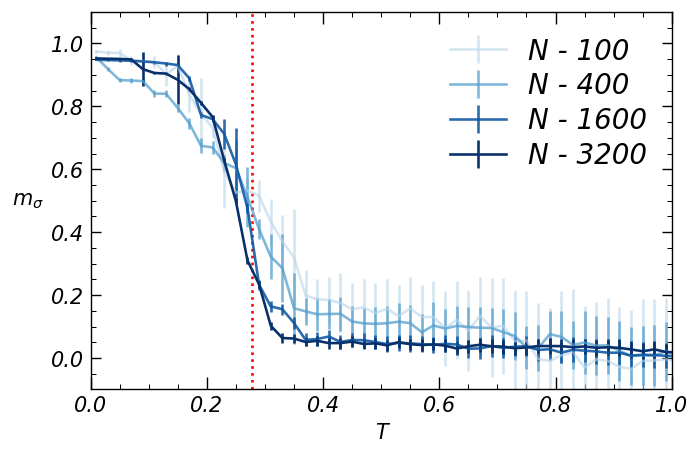}
    \caption{Finite size scaling MC simulations of a symmetric ($\gamma=1$) BAM network at a load of $\alpha=0.1$ with equal dataset entropy per layer: $\rho_s= \bar \rho_s$. The number of examples is $M=10$ in each simulation reported. For each plot, four curves are displayed, in different blue tonalities (as function of different values of size $N=100, 400, 1600, 3200$), as indicated in the legends. Each plot is relative a different set of MC simulations, differing solely for the dataset entropy: in particular, from left to right, $\rho_s=0, 0.05, 0.1$. The dotted vertical red lines depicts the theoretical critical temperature as predicted by Statistical Mechanics.}
    \label{Fig:trackstraj}
\end{figure}

These equations have been investigated numerically and results are presented in Fig. \ref{Fig:PD_sup}.
In Figure \ref{Fig:PD_sup} (left panel) we show the retrieval region in the $\alpha,\ \beta^{-1}$ plane, at different values of the dataset entropy (that is kept identical in the two datasets provided to the two layers of the BAM) as reported by the legend, further in the right panel we confirm that --also moving from storing to learning-- the maximal retrieval region is obtained by keeping the ratio among the two layers of $O(1)$ (this is intuitive as this condition maximizes the amount of synapses -where the information is stored- in the network) \cite{BAM}. 
\newline
In Figure \ref{Fig:trackstraj} we provide finite size scaling Monte Carlo runs to depict the transition toward the retrieval region in the noise $\beta^{-1}$: we show solely the magnetization of one layer (the other being identical as these plots are worked out for symmetric layers) as the noise (i.e., the temperature) is tuned. 

\subsection{Ground-state analysis ($\beta \to \infty$ limit)}\label{Sezione3.3}
In the present theory there are two sources of fast noise, the one that lies in the dataset -ruled by $\rho_s$ and $\bar{\rho_s}$- and the one affecting the machine -ruled by $\beta$- and it is interesting to deepen in particular the network's capabilities in the ground-state, that is when $\beta \to \infty$ as those are expected to be maximal\footnote{Indeed, while the role of $r, \bar{r}$ (and thus of $\rho$ and $\bar{\rho}$) is in some sense unavoidable because data come jointly with their noise, the network can work in the ground-state and, depending on the usage of the machine, one can tune $\beta$ accordingly, e.g. if we ask the BAM to classify it is better to remove noise in the network, while if we use it as a generative model, then $\beta$ plays a crucial role in blurring an archetype toward the generation of a new example.}.

The critical load of the model can be derived by directly taking the limit $T=\beta^{-1}\to 0$ in the self-consistency equations (\ref{eq:ns_sup})-(\ref{eq:pbar_sup}), and by requiring the quantities $\chi = \beta(1-q)$ and $\overline \chi = \beta(1-\bar q)$ to remain finite in the same limit. The equations read:
\begin{align}
    &n_\sigma = \frac{1}{\sqrt{(1+\rho_s)(1+\bar \rho_s)}} \frac{m_\sigma + \chi m_\tau \gamma^{-1}\frac{\rho_s}{\sqrt{(1+\rho_s)(1+\bar \rho_s)}}}{1-\chi \overline \chi \frac{\rho_s\bar \rho_s}{(1+\rho_s)(1+\bar \rho_s)}},\label{eq:ns_sup_zeroT}\\
    &n_\tau = \frac{1}{\sqrt{(1+\rho_s)(1+\bar \rho_s)}} \frac{m_\tau + \overline \chi m_\sigma \gamma\frac{\bar \rho_s}{\sqrt{(1+\rho_s)(1+\bar \rho_s)}}}{1-\chi \overline \chi \frac{\rho_s\bar \rho_s}{(1+\rho_s)(1+\bar \rho_s)}},\label{eq:nt_sup_zeroT}\\
    &m_\sigma = \erf(y),\label{eq:ms_sup_zeroT}\\
    &m_\tau = \erf(\bar y),\label{eq:mt_sup_zeroT}\\
    &y = \frac{n_\tau \gamma^{-1}}{\sqrt{2\alpha \frac{1+\overline\chi^2}{(1-\chi \overline \chi)^2}+2\rho_s(n_\tau \gamma^{-1})^2}},\label{eq:y}\\
    &\bar y = \frac{n_\sigma \gamma}{\sqrt{2\bar \alpha \frac{1+\chi^2}{(1-\chi \overline \chi)^2}+2\bar \rho_s(n_\sigma \gamma)^2}}\label{eq:ybar}.
\end{align}

\begin{figure}[!ht]    
    \centering
    \includegraphics[width=7.5cm]{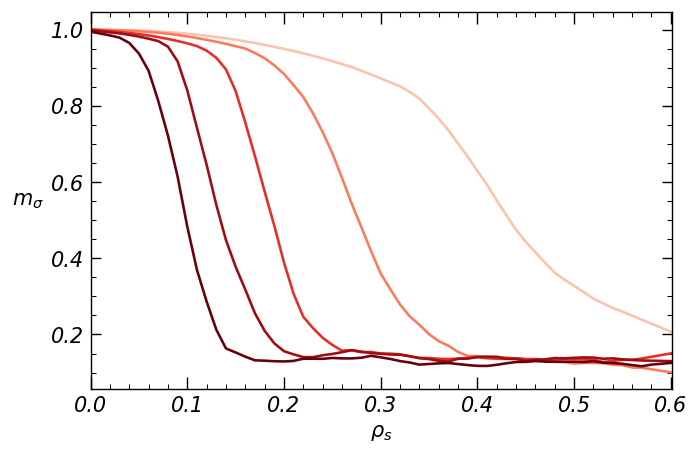}
    \includegraphics[width=7.25cm]{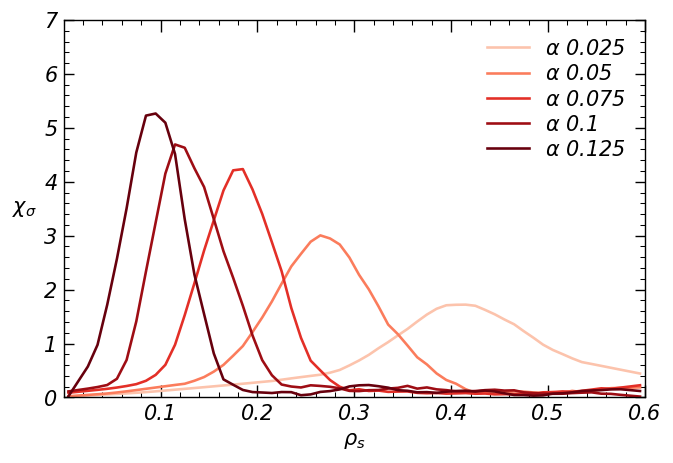}
    \caption{Left: the mattis magnetization is displayed as a function of $\rho_s$ for various values of the load $\alpha = \bar \alpha$ at $\beta^{-1}=0.1$. Right: The susceptibility $\chi_\sigma = \frac{\partial m_\sigma}{\partial \rho_s}$ is displayed for different values of the load $\alpha = \bar \alpha$. The model is symmetric, \emph{i.e.} $\gamma = 1$ and also the noise $\rho_s= \bar \rho_s$ is the same in both layers.}
    \label{Fig:info-transition}
\end{figure}

From these equations we can inspect a computational transition line that is not achievable in the standard storing theory (nor in AGS theory neither in the Kosko theory of the BAM), namely we deepen the behavior of the magnetizzation $m_{\sigma}$ and of its susceptibility $\chi_{\sigma}$ as the entropy of the dataset is made to vary (again we present solely one magnetization and discard $m_{\tau}$ as these plots are achieved for the symmetric BAM, where the two layers behave identically): as we show in Figure \ref{Fig:info-transition}, by increasing the information supplied to the network (that is, by lowering the entropy related to the datasets that the network experiences)  the network eventually raises its magnetization proving itself able to retrieve by reverberation. We also note that, the higher the load $\alpha$ the lower the entropy of the dataset required by the network to perform correctly, as intuitive.

\bigskip

\begin{figure}[!ht]    
    \centering
    \includegraphics[width=5cm]{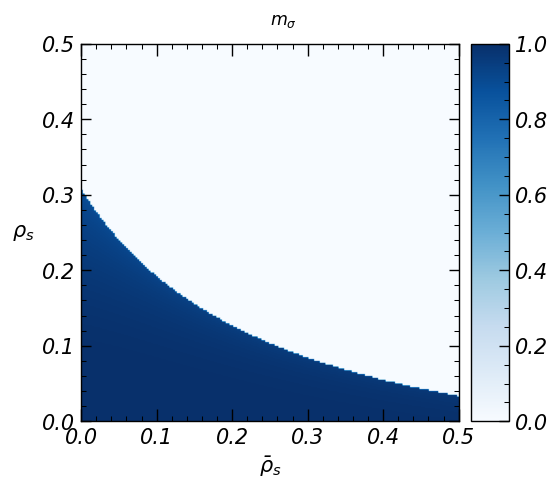}
    \includegraphics[width=5cm]{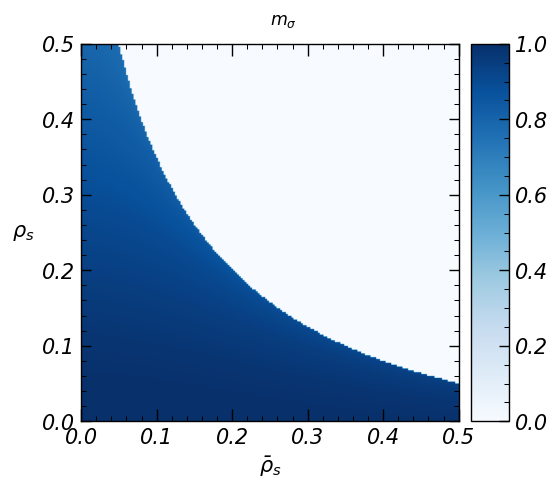}
    \includegraphics[width=5cm]{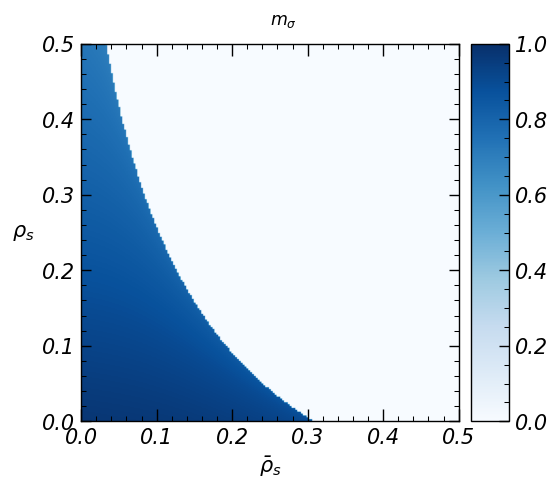}
    \caption{The magnetization $m_\sigma$ is shown in tones of blue, as a function of the noise entropy $\rho_s$ and $\overline \rho_s$, in three configurations, from left to right: $\lambda=0.02$-$\gamma=0.1$, $\lambda=0.07$-$\gamma=1$ and $\lambda=0.02$-$\gamma=10$.}
    \label{Fig:2splot1}
    \includegraphics[width=5cm]{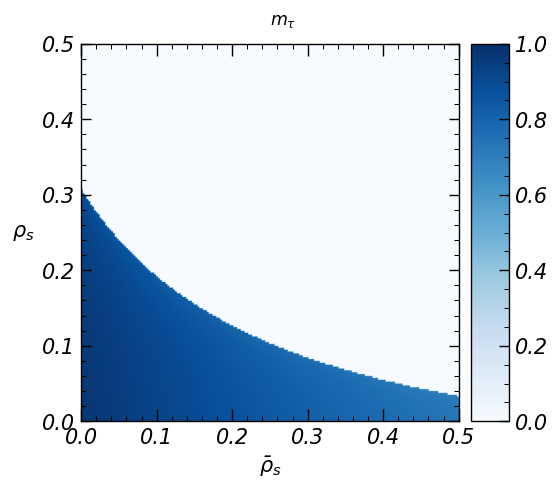}
    \includegraphics[width=5cm]{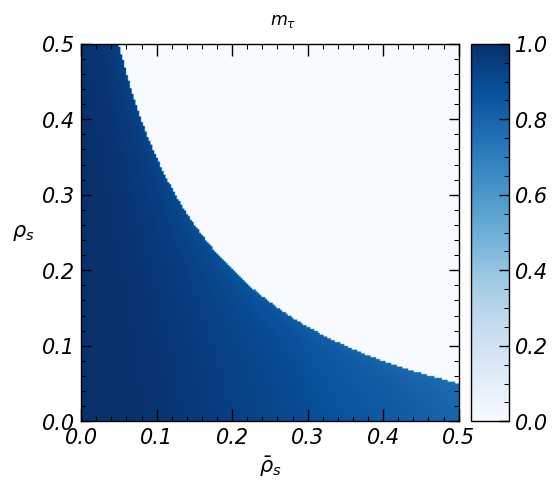}
    \includegraphics[width=5cm]{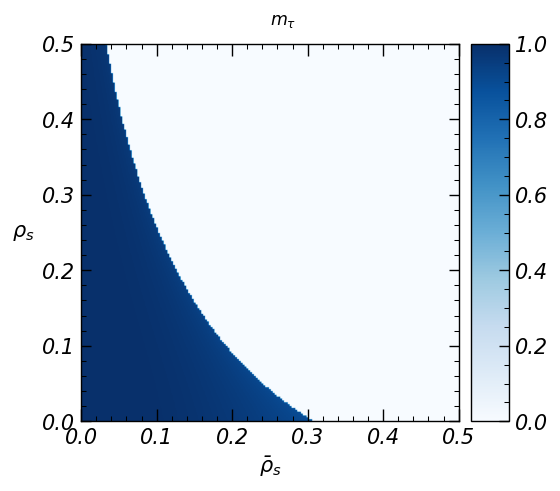}
    \caption{The magnetization $m_\tau$ is shown in tones of blue, as a function of the noise entropy $\rho_s$ and $\overline \rho_s$, in three configurations, from left to right: $\lambda=0.02$-$\gamma=0.1$, $\lambda=0.07$-$\gamma=1$ and $\lambda=0.02$-$\gamma=10$.}
    \label{Fig:2splot2}
\end{figure}

We can further appreciate how coupled retrieval takes place via reverberation in the BAM and how the network can take advantages of this by a glance at Figures  \ref{Fig:2splot1} (related to the layer $\{\sigma\}$) and \ref{Fig:2splot2} (related to the layer $\{\tau\}$): the three panels per raw show three different BAMs (one per column), namely strongly asymmetric with a small layer $\{\sigma\}$ (left, $\gamma=0.1$), symmetric (center, $\gamma=1$) and strongly asymmetric with a small layer $\{\tau\}$ (left, $\gamma=10$).  
\newline
Let us focus on the symmetric case for the sake of simplicity (hence we look at the central plots in Figures \ref{Fig:2splot1} and \ref{Fig:2splot2}): even if one of the two visible layers, say $\tau$, is provided with a very noisy dataset (hence the network is forced in the $\bar{\rho}_s \gg 0$ region of these plots), yet it can take advantage of a proper pattern recognition in the other layer, that must be provided with an informative dataset (i.e. the network must approach the $\rho_s\to 0$ region of these plots) to guarantee the correct reverberation among the layers until convergence to both the correct archetypes. 
\newline
This can easily be understood also from the dual perspective of the two coupled RBMs interacting via their grand-mother cells among different hidden layers: indeed, once a RBM succeeded to correctly retrieve the related archetype, say $\xi^{\nu}$, it has the corresponding hidden grand-mother neuron $z_{\nu}$ firing (that is $z_{\nu}>>1$); this -in turn- drives its conjugated  grand-mother neuron $z_{\nu}^{\dag}$ in the other RBM to fire too (that is $z^{\dag}_{\nu}>>1$), guiding the latter toward the correct retrieval of $\bar{\xi}^{\nu}$: see Figure \ref{Duality}.
\newpage
\section{Replica symmetric theory of Unsupervised Learning via Reverberation}\label{SezioneQuattro}

In this Section we repeat the calculations provided in the previous Section but focusing on Unsupervised protocol rather than the Supervised one. We do not report in detail all the passages, rather we outline just the main steps (as calculations are much the same as before).
\newline
After decomposing the Hamiltonian, eq. (\ref{eq:unsup}), in a signal containing the archetype to be learnt and retrieved ($\mu=1$) and the noise made of the background (i.e. all the information regarding $\mu>1$), we write the partition function as
\begin{align} 
    Z_{N,\bar{N},M,\bar{M}}=\sum_{\sigma,\tau}\exp\lr{ \frac{\beta L}{M\sqrt{R\overline R}} \sum_a n^{1a}_\sigma n^{1a}_\tau + \frac{\beta}{LM\sqrt{R\overline R}} \sum_{\mu>1} \sum_{ik} \sum_a \eta^{\mu a}_i \bar \eta^{\mu a}_k \sigma_i \tau_k }.
    \label{eq:unsupZ}
\end{align}
In the limit of large number of examples $M\gg1$, we use the CLT to rewrite $\sum_a \eta^{\mu a}_i \bar \eta^{\mu a}_k $ as follows:
\begin{align}
    &\sum_a \eta^{\mu a}_i \bar \eta^{\mu a}_k \sim r \bar r \sqrt{1+\rho} \: z^\mu_{ik}, \:\: z^\mu_{ik}\sim \mathcal N (0,1),
\end{align}
where $\rho$ is the parameter defined in eq. (\ref{eq:rho_unsup}); this allows us to rewrite the partition function as:
\begin{align}
    Z_{N,\bar{N},M,\bar{M}}=\sum_{\sigma,\tau}\exp\lr{ \frac{\beta}{M L\sqrt{R\overline R}} 
    \sum_a \sum_{i,k} \eta^{1 a}_i \bar \eta^{1 a}_k \sigma_i \tau_k
    +
    \frac{\beta}{L} \frac{\sqrt{1+\rho}}{\sqrt{(1+\rho_s)(1+\bar \rho_s)}} \sum_{\mu>1} \sum_{ik} z^\mu_{ik}\sigma_i \tau_k }.
\end{align}
Following the Guerra's method exposed in the previous section at work with the supervised counterpart we arrive to an expression for the quenched statistical pressure in the \emph{RS} approximation, $A_{RS}(\beta)$. The details of the computation are given in Appendix \ref{unsupAppendix}. The $A_{RS}(\beta)$ pressure reads:
\begin{align}\label{eq:ARSunsup}
    A_{RS}(\beta) =& - \beta \sqrt{(1+\rho_s)(1+\bar \rho_s)} \:n_\sigma n_\tau + \frac{\lambda \beta^2}{2} \frac{1+\rho}{(1+\rho_s)(1+\bar \rho_s)} (1-q)(1-\bar q)+\nonumber\\
    +&\gamma \mathbf E_{\eta} \int Dx \ln \cosh \lr{\beta\sqrt{\frac{1+\rho}{(1+\rho_s)(1+\bar \rho_s)} \alpha \bar q}\: x + \beta\frac{n_\tau \gamma^{-1}}{r} \lr{\frac{1}{M}\sum_a \eta^{1a}} }+ \nonumber\\
    +&\gamma^{-1} \mathbf E_{\bar \eta} \int Dx \ln \cosh \lr{\beta\sqrt{\frac{1+\rho}{(1+\rho_s)(1+\bar \rho_s)} \bar \alpha q}\: x + \beta\frac{n_\sigma \gamma}{\bar r} \lr{\frac{1}{M}\sum_a \bar \eta^{1a}} }
\end{align}
Note that we have used: $\lambda \gamma^{-1} = \alpha$ and $\lambda \gamma = \bar \alpha$.\\
The self-consistence equations are:
\begin{align}
    &n_\sigma = \mathbf E_\eta \lr{ \frac{1}{M} \sum_a \eta^a} \int Dx \tanh \lr{\beta\sqrt{\frac{1+\rho}{(1+\rho_s)(1+\bar \rho_s)} \alpha \bar q}\: x + \beta\frac{n_\tau \gamma^{-1}}{r} \lr{\frac{1}{M}\sum_a \eta^{1a}} }\\
    &n_\tau = \mathbf E_{\bar\eta} \lr{ \frac{1}{M} \sum_a \bar \eta^a} \int Dx \tanh \beta\sqrt{\frac{1+\rho}{(1+\rho_s)(1+\bar \rho_s)} \bar \alpha q}\: x + \beta\frac{n_\sigma \gamma}{\bar r} \lr{\frac{1}{M}\sum_a \bar \eta^{1a}}\\
    &m_\sigma = \mathbf E_\eta\int Dx \tanh \lr{\beta\sqrt{\frac{1+\rho}{(1+\rho_s)(1+\bar \rho_s)} \alpha \bar q}\: x + \beta\frac{n_\tau \gamma^{-1}}{r} \lr{\frac{1}{M}\sum_a \eta^{1a}} }\\
    &m_\tau = \mathbf E_{\bar\eta}\int Dx \tanh \beta\sqrt{\frac{1+\rho}{(1+\rho_s)(1+\bar \rho_s)} \bar \alpha q}\: x + \beta\frac{n_\sigma \gamma}{\bar r} \lr{\frac{1}{M}\sum_a \bar \eta^{1a}}\\
    &q = \mathbf E_\eta\int Dx \tanh^2 \lr{\beta\sqrt{\frac{1+\rho}{(1+\rho_s)(1+\bar \rho_s)} \alpha \bar q}\: x + \beta\frac{n_\tau \gamma^{-1}}{r} \lr{\frac{1}{M}\sum_a \eta^{1a}} }\\
    &\bar q = \mathbf E_{\bar\eta}\int Dx \tanh^2 \beta\sqrt{\frac{1+\rho}{(1+\rho_s)(1+\bar \rho_s)} \bar \alpha q}\: x + \beta\frac{n_\sigma \gamma}{\bar r} \lr{\frac{1}{M}\sum_a \bar \eta^{1a}}.
\end{align}
\begin{figure}
    \centering
    \includegraphics[width=5cm]{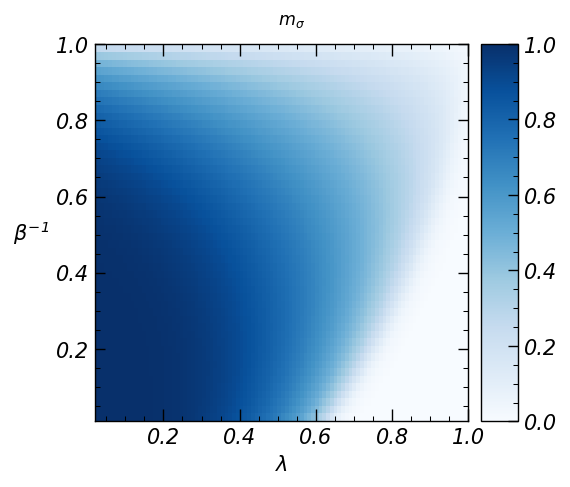}
    \includegraphics[width=5cm]{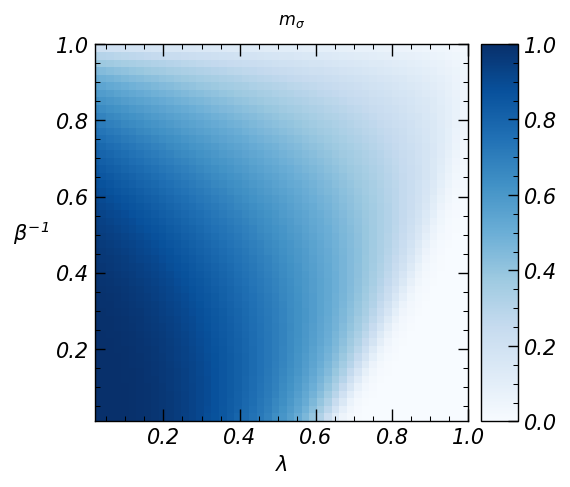}
    \includegraphics[width=5cm]{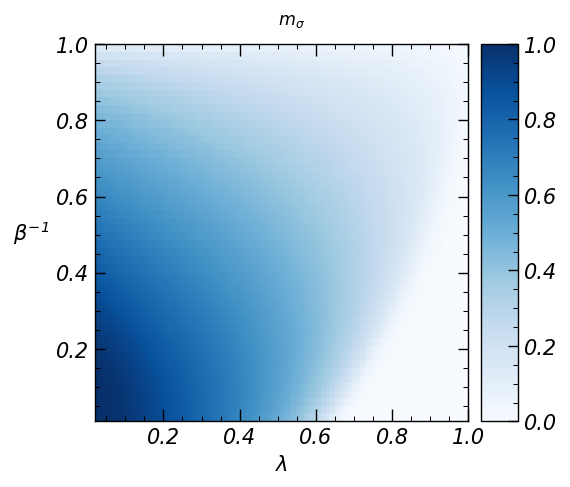}
    \includegraphics[width=5cm]{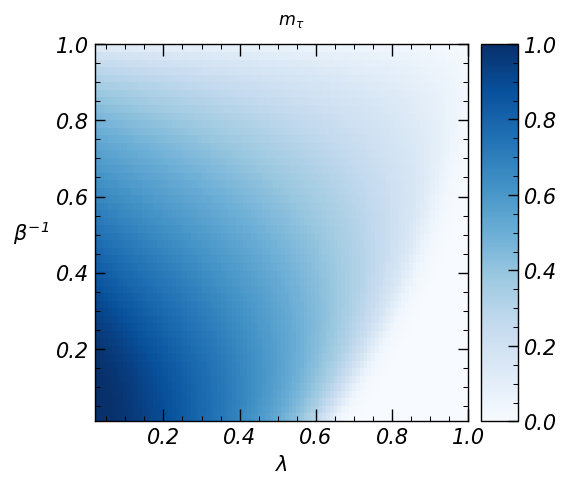}
    \includegraphics[width=5cm]{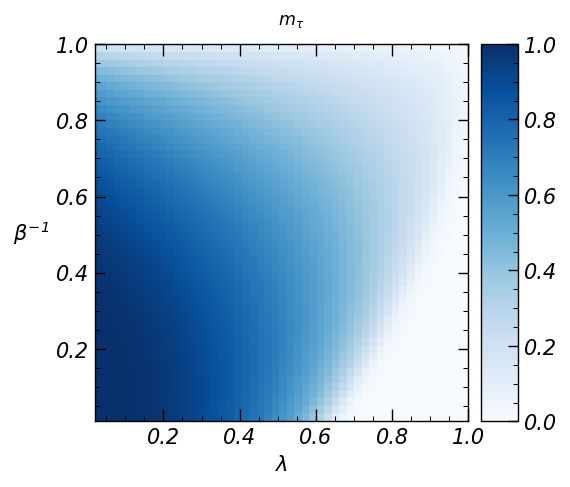}
    \includegraphics[width=5cm]{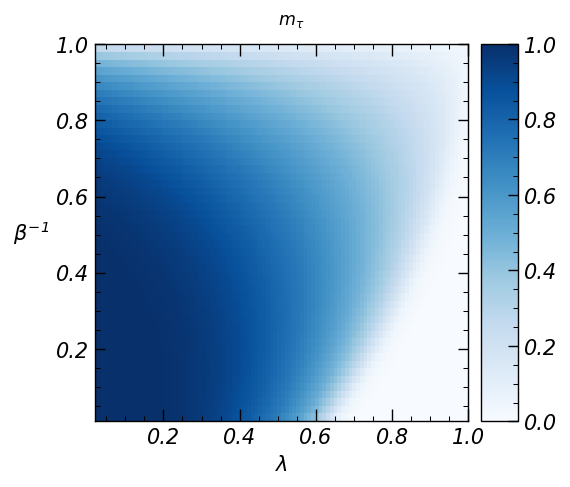}
    \caption{Numerical solutions of the self-consistency equations \ref{eq:unsup_ns}-\ref{eq:unsup_qt}, for $\gamma=0.5$ (\emph{left}), $\gamma=1.0$ (\emph{center}), $\gamma=2.0$ (\emph{right}). The dataset entropy is given by $r = \bar r = 0.9$ and $M=100$.}
    \label{fig:unsup_rho002346}
\end{figure}

\begin{figure}
    \centering
    \includegraphics[width=5cm]{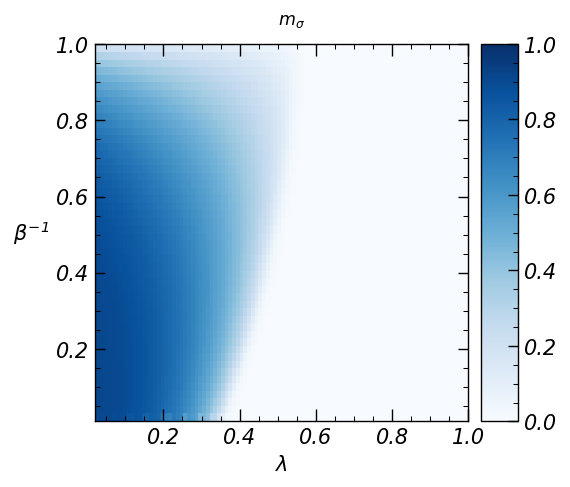}
    \includegraphics[width=5cm]{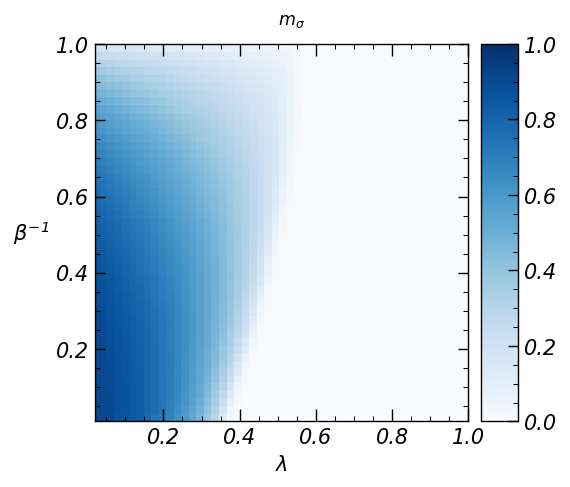}
    \includegraphics[width=5cm]{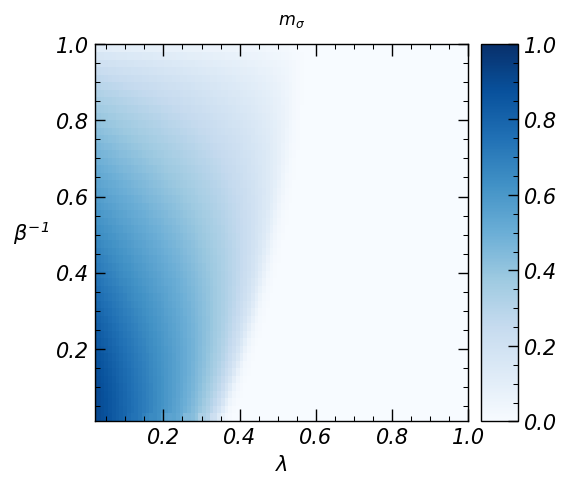}
    \includegraphics[width=5cm]{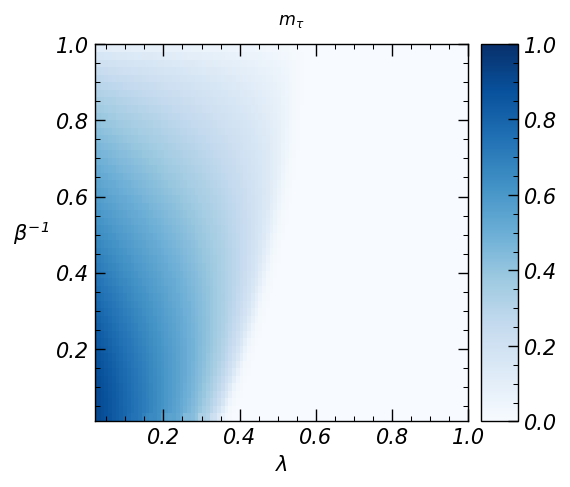}
    \includegraphics[width=5cm]{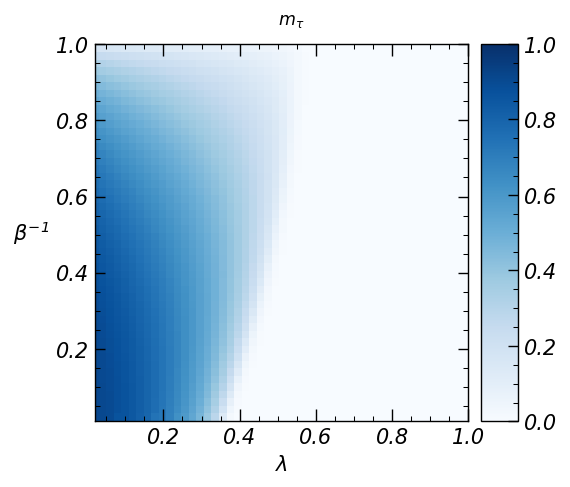}
    \includegraphics[width=5cm]{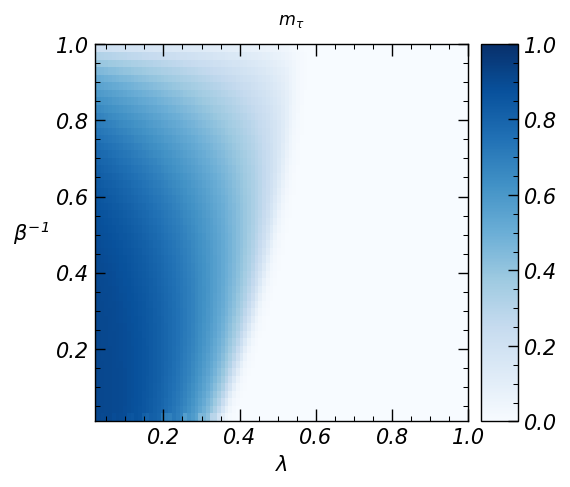}
    \caption{Numerical solution of the self-consistency equations \ref{eq:unsup_ns}-\ref{eq:unsup_qt}, for $\gamma=0.5$ (\emph{left}), $\gamma=1.0$ (\emph{center}), $\gamma=2.0$ (\emph{right}). The dataset entropy is given by $r = \bar r = 0.3$ and $M=100$.}
    \label{fig:unsup_rho1011}
\end{figure}
\subsection{Big data regime}
In the $M\to\infty$ limit we can use the CLT to express the algebraic average of the examples as we did in the supervised setting; see equations (\ref{eq:fluctuations_1},\ref{eq:fluctuations_2}). After a few rearrangements, the self-consistency equations are:
\begin{align}
    &n_\sigma = \frac{1}{\sqrt{(1+\rho_s)(1+\bar \rho_s)}} \frac{m_\sigma + \beta (1-q) m_\tau \gamma^{-1}\frac{\rho_s}{\sqrt{(1+\rho_s)(1+\bar \rho_s)}}}{1-\beta^2(1-q)(1-\bar q)\frac{\rho_s\bar \rho_s}{(1+\rho_s)(1+\bar \rho_s)}}\label{eq:unsup_ns}\\
    &n_\tau = \frac{1}{\sqrt{(1+\rho_s)(1+\bar \rho_s)}} \frac{m_\tau + \beta (1-\bar q) m_\sigma \gamma\frac{\bar \rho_s}{\sqrt{(1+\rho_s)(1+\bar \rho_s)}}}{1-\beta^2(1-q)(1-\bar q)\frac{\rho_s\bar \rho_s}{(1+\rho_s)(1+\bar \rho_s)}}\label{eq:unsup_nt}\\
    &m_\sigma = \int Dx \tanh g_1\label{eq:unsup_ms}\\
    &m_\tau = \int Dx \tanh g_2\label{eq:unsup_mt}\\
    &q = \int Dx \tanh^2 g_1\label{eq:unsup_qs}\\
    &\bar q = \int Dx \tanh^2 g_2,\label{eq:unsup_qt}
\end{align}
where we have introduced the functions $g_1, g_2$
\begin{align}
    &g_1 = \beta \sqrt{\frac{ 1 + \rho}{(1+\rho_s)(1+\bar \rho_s)} \alpha \bar q + \rho_s (n_\tau \gamma^{-1})^2}\: x + \beta n_\tau \gamma^{-1}\\
    &g_2 = \beta \sqrt{\frac{ 1 + \rho}{(1+\rho_s)(1+\bar \rho_s)} \bar \alpha q + \bar \rho_s (n_\sigma \gamma)^2}\: x + \beta n_\sigma \gamma.
\end{align}
Graphical outcomes of these solutions is provided in Figures \ref{fig:unsup_rho002346} and \ref{fig:unsup_rho1011}

\section{Signal to Noise analysis}\label{SezioneCinque}
Once provided an exhaustive statistical mechanical picture we inspect the quality of the  ground state solutions by the signal-to-noise analysis to have an independent confirm of the previous findings (as these have been derived under the replica symmetric approximation). The core idea of this technique is rather simple: once assumed $(\sigma, \tau)$ to be a stable configuration corresponding to a stored archetype, the stability request for this configuration simply reads as $\sigma_i h_i \geq 0$ for all $i \in (1,...,N)$ (for one layer) and $\tau_k h_k \geq 0$ for all $k \in (1,...,\bar{N})$ (for the other layer), where we called $h$ the post-synaptic fields acting on these neurons.
\newline
Hereafter we derive the conditions by which the given neuron $\sigma_i$ (and similarly $\tau_i$), subjected to the internal field $h_i$, is aligned to archetype $\xi^1$ (and, similarly, $h_k$ for $\bar{\xi}^1$) that we want the network to retrieve, splitting the cases of supervised and unsupervised protocols in the next two subsections.\\
As we assume that the configuration $(\sigma, \tau)$ is a retrieval one, the analysis  may be implemented following the one-step update prescription of Hinton for fast MC sampling \cite{early}. We begin by specifying the dynamics of the model at $\beta\to\infty$, as expressed through the update equations at time $t+1$, keeping in mind that the goal is to evaluate the state of the neuron $\sigma$ at time $t=1$ starting from the initial configuration $\sigma^{(t=0)}=\xi^1, \tau^{(t=0)}=\bar\xi^1$:
\begin{align}
    &\sigma^{(t+1)} = \sign(h^{(t)}_i \sigma^{(t)}_i),\label{eq:updates}\\
    &\tau^{(t+1)} = \sign(\overline h^{(t)}_k \tau^{(t)}_k).
\end{align}
The neuronal fields $h_i$ and $\bar h_k$ are the internal fields felt by the neurons $\sigma_i$ and $\tau_k$; they are derived from the definition of the Hamiltonian of the model. In particular, in the the supervised setting from eq. (\ref{eq:sup}) follows:
\begin{align}
    &h^{sup}_i = \frac{1}{LM^2 \sqrt{R \overline R}} \sum_{\mu=1^K} \sum_{k=1}^{\overline N} \left(\sum_{a=1}^{M} \eta^{\mu a}_i\right) \left(\sum_{b=1^M} \overline \eta^{\mu b}_k\right) \tau_k,\\
    &\overline h^{sup}_k = \frac{1}{LM^2 \sqrt{R \overline R}} \sum_{\mu=1}^{K} \sum_{i=1}^{N} \left(\sum_{a=1}^{M} \eta^{\mu a}_i\right) \left(\sum_{b=1}^{M} \overline \eta^{\mu b}_k\right) \sigma_i;
\end{align}
while in the unsupervised case, from eq. (\ref{eq:unsup}) we have
\begin{align}
    &h^{unsup}_i = \frac{1}{L} \sum_{\mu=1}^{K} \sum_{k=1}^{\overline N} \sum_{a=1}^{M} \eta^{\mu a}_i \overline \eta^{\mu a}_k \tau_k,\\
    &\overline h^{unsup}_k = \frac{1}{L} \sum_{\mu=1}^{K} \sum_{i=1}^{N} \sum_{a=1}^{M} \eta^{\mu a}_i \overline \eta^{\mu a}_k \sigma_i.
\end{align}
After deriving the state of the network at time $t=1$, $\sigma^{(t=1)},\tau^{(t=1)}$, the overlap between this state and the patterns $\xi^1, \overline \xi^1$ is given by the Mattis magnetizations $m^1_\sigma,m^1_\tau$ as follows:
\begin{align}
    &m^1_\sigma = \frac{1}{N} \sum_i \xi^1_i \sigma^{(t=1)}\\
    &m^1_\tau = \frac{1}{\overline N} \sum_k \overline \xi^1_k \tau^{(t=1)}.
\end{align}
These quantities can be evaluated in the large number of examples limit $M\to\infty$ (big data scenario), by replacing the algebraic average $\langle . \rangle = \frac{1}{N} \sum_i (.)$ with a Gaussian integral over a stochastic variable $z\sim \mathcal{N}(0,1)$, by virtue of the CLT, as follows:
\begin{align}
    \frac{1}{N} \sum_i \xi^1_i \sigma^{(t=1)} = \int Dz \: \xi^1\sign(\mu + s z)\label{eq:gaussianavg}
\end{align}
where we have expressed $\sigma^{(t=1)}$ through its update rule eq. (\ref{eq:updates}), and introduced the mean and variance $\mu,s$ of the argument of the sign function, given by:
\begin{align}
    &\mu = \mathbf E \left[ h^{(t=0)} \xi^1\right],\\
    &s = \mathbf E \left[\left( h^{(t=0)} \xi^1\right)^2\right] - \mu^2,\\
\end{align}
where now formally $\xi^1$ is a binary scalar variable, $\xi^1 = \pm 1$,
and $\mathbf E = \mathbf E_{\xi,\overline \xi, \chi, \overline \chi}$. Notice that, by simply rescaling $z \to \xi^1 z$, we can rewrite eq. (\ref{eq:gaussianavg}) as
\begin{align}
    m_\sigma^1 = \int Dz \: \sign(\mu + \sqrt s z) = \erf \left(\frac{\mu}{\sqrt{2 s}}\right)\label{eq:gaussianavg2}.
\end{align}

In the following we outline the computation of $m_\sigma^1$ through eq. (\ref{eq:gaussianavg2}), in the supervised and unsupervised cases. Equivalently, the computation can be similarly done for deriving $m_\tau^1$. 

\subsection{Signal-to-noise for the Supervised setting}\label{Sezione5.1}
\begin{figure}[!ht]    
    \centering
    \includegraphics[width=7.5cm]{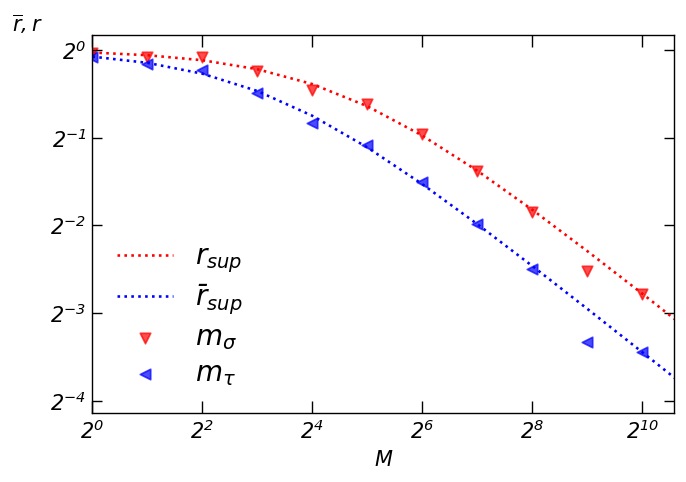}
    \includegraphics[width=7.5cm]{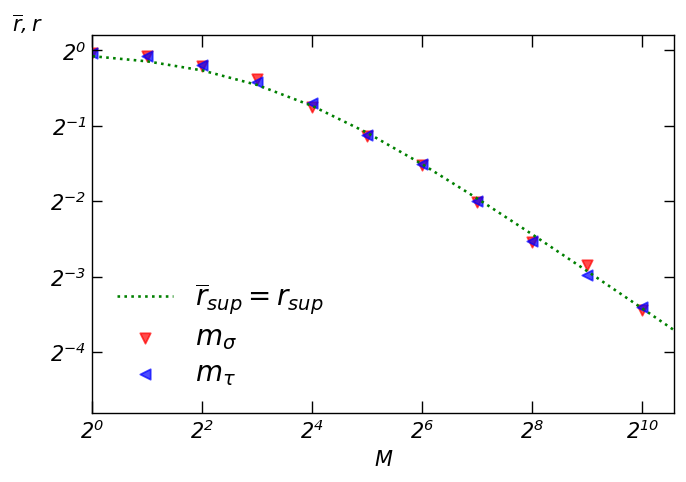}
    \caption{\label{S2N-Sup} The theoretical predictions derived in Equations (\ref{eq:supervised_ThrForLearning_1}),\ (\ref{eq:supervised_ThrForLearning_2}) are compared to simulated data for two different configurations of the network in the supervised scenario: asymmetric (\emph{left} - $N=200, \overline N=100$) and symmetric (\emph{right} - $N=200, \overline N=200$). The number of stored patterns is $K=12$. The plots are obtained by defining a fixed threshold for $m_\sigma$ and $m_\tau$, resulting in a condition on the level of dataset entropy $r$ and $\bar r$ as a function of $M$, as shown in the figures.}
\end{figure}

Firstly we start by computing $\mu$. It is given by
\begin{align}
    \mu = \frac{1}{LM^2 \sqrt{R \overline R}} \mathbf E\left[ \sum_{\mu=1}^{K} \sum_{k=1}^{\overline N} \left(\sum_{a=1}^{M} \eta^{\mu a} \right) \left(\sum_{b=1}^{M} \overline \eta^{\mu b}_k\right)  \overline \xi_k^1 \xi^1 \right].
\end{align}
It can be divided in two terms, the first for $\mu=1$ and the second for $\mu>1$:
\begin{align}
    \mu = \frac{1}{LM^2 \sqrt{R \overline R}} \mathbf E\left[ \sum_{k} \sum_{a,b} \chi^{1 a}  \overline \chi^{1 b}_k \right] + \frac{1}{LM^2 \sqrt{R \overline R}} \mathbf E\left[ \sum_{\mu>1} \sum_{k} \sum_{a,b} \overline \xi^\mu_k \overline \xi^1_k\chi^{\mu a}  \overline \chi^{\mu b}_k \xi^1 \xi^\mu \right].
\end{align}
The last term is zero given that $\mathbf E \left[ \overline \xi_k^\mu \xi_k^\nu \right] = \delta^{\mu\nu}$ and $\mu\neq 1$ in this case.\\
Thus we have
\begin{align}
    \mu = \frac{1}{LM^2 \sqrt{R \overline R}} \sum_{k} \sum_{a,b} \mathbf E_{\chi} \left[ \chi^{1 a}  \right] \mathbf E_{\overline \chi}\left[\overline \chi^{1 b}_k \right] = \frac{\overline N r \bar r}{L\sqrt{R \overline R}} = \frac{\gamma^{-1}}{\sqrt{(1+\rho_s)(1+\bar\rho_s)}},
\end{align}
given the definition of $R$, $\overline R$, $\rho_s$ and $\bar \rho_s$.\\
In order to compute $s$, we need to derive $\mathbf E \left[\left( h^{(t=0)} \xi^1\right)^2\right]$:
\begin{align}
    \mathbf E \left[\left( h^{(t=0)} \xi^1\right)^2\right] = \frac{1}{L^2M^4 R \overline R} \mathbf E\left[ \left(\sum_{\mu=1}^{K} \sum_{k=1}^{\overline N} \left(\sum_{a=1}^{M} \eta^{\mu a} \right) \left(\sum_{b=1}^{M} \overline \eta^{\mu b}_k\right) \overline \xi_k^1 \xi^1 \right)^2\right],
\end{align}
which can be again decomposed in two contributions, $\mu=1$ and $\mu>1$:
\begin{align}
    \frac{1}{L^2M^4 R \overline R} \mathbf E\left[ \left( \sum_{k} \sum_{a,b} \chi^{1 a} \overline \chi^{1 b}_k\right)^2\right] + \frac{1}{L^2M^4 R \overline R} \mathbf E\left[ \left( \sum_{\mu>1}\sum_{k} \sum_{a,b} \chi^{\mu a} \bar \chi^{\mu b}_k \: \bar \xi_k^\mu \bar \xi_k^1 \xi^\mu \xi^1\right)^2\right].
\end{align}
The first term can be computed as follows:
\begin{align}
    \frac{1}{L^2M^4 R \overline R} \sum_{k,l} \sum_{a,b,c,d} \mathbf E\left[ \chi^{1 a} \chi^{1 c} \overline \chi^{1 b}_k \overline \chi^{1 d}_l \right] = \frac{\overline N^2 M^2 \bar r^2}{L^2M^4 R \overline R} (M^2 r^2 + M(1-r^2)) = \frac{\gamma^{-2}}{1+\bar \rho_s}.
\end{align}
Carefully inspecting the second term, it can be seen that only the term of order $\mathcal O (\overline N K)$ survives; it simply reads:
\begin{align}
    \frac{1}{L^2M^4 R \overline R} \sum_{\mu>1}\sum_{k} \sum_{a,b,c,d} \mathbf E\left[ \chi^{\mu a} \chi^{\mu c} \bar \chi^{\mu b}_k \bar \chi^{\mu d}_k \right] = \frac{K \overline N M^4 r^2 \bar r^2 (1+\rho_s)(1+\bar \rho_s)}{L^2M^4 R \overline R} =\alpha .
\end{align}
Collecting all the terms, we arrive at the following expression for $s$:
\begin{align}
    s = \alpha + \frac{\gamma^{-2}}{1+\bar \rho_s} - \frac{\gamma^{-2}}{(1+\rho_s)(1+\bar\rho_s)} = \alpha + \frac{\gamma^{-2} \rho_s}{(1+\rho_s)(1+\bar\rho_s)}.
\end{align}
Hence, $m^1_\sigma$ finally reads:
\begin{align}
    m^1_\sigma = \erf\left( \frac{1}{\sqrt{2\left(\rho_s+ \bar \alpha (1+\rho_s)(1+\bar\rho_s) \right)} }\right).
\end{align}
The stability condition for the retrieval by the layer $\sigma$ of the pattern $\xi^1$ is then
\begin{align} 
    \frac{1-r^2}{Mr^2} + \bar \alpha \left(1+\frac{1-r^2}{Mr^2}\right)\left(1+\frac{1-\bar r^2}{M\bar r^2}\right)< 1.
    \label{eq:supervised_ThrForLearning_1}
\end{align}
The stability condition for the retrieval by the layer $\tau$ of the pattern $\bar \xi^1$ is then
\begin{align} 
    \frac{1-\bar r^2}{M \bar r^2} + \alpha \left(1+\frac{1-r^2}{M r^2}\right)\left(1+\frac{1-\bar r^2}{M\bar r^2}\right)< 1.
    \label{eq:supervised_ThrForLearning_2}
\end{align}
As shown in Figure \ref{S2N-Sup}, the above conditions -eq.s (\ref{eq:supervised_ThrForLearning_1}) and (\ref{eq:supervised_ThrForLearning_2})- provide a recipe to guarantee, before starting to train the network, if we have enough information to provide to it to ensure the correct learning, storage and successive retrieval of the archetypes over both the layers: in Figure \ref{Esempi} we report the growth of the magnetizations of the two archetypes under retrieval as more and more examples are supplied to the network. 

\subsection{Signal-to-noise for the Unsupervised setting}\label{Sezione5.2}
\begin{figure}[!ht]    
    \centering
    \includegraphics[width=7.5cm]{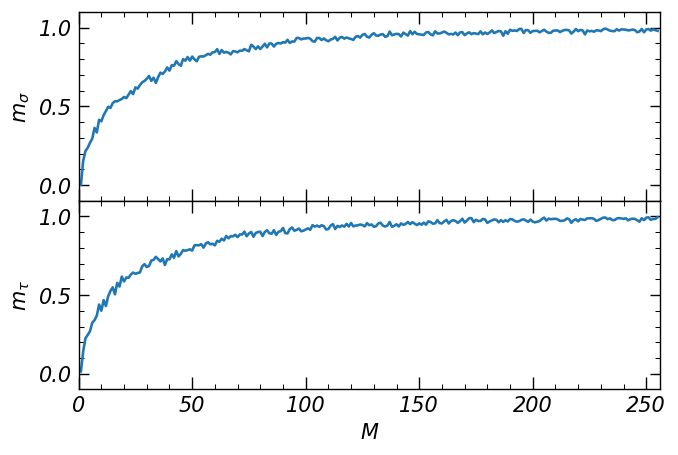}
    \includegraphics[width=7.5cm]{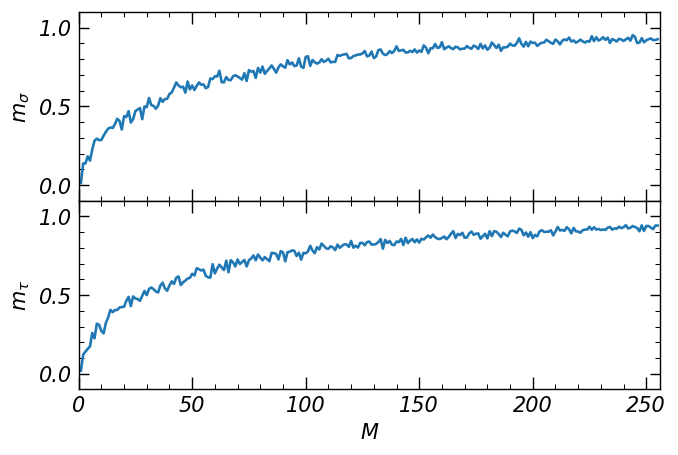}
    \caption{\label{Esempi} The retrieval quality of the models trained by the supervised (\emph{left}) and unsupervised (\emph{right}) protocols are shown in terms of the magnetizations $m_\sigma$ and $m_\tau$, as a function of the number of examples $M$. The models have been taken symmetric with $N=\overline N = 200$ and a number of patterns $K=12$; $r=\bar r = 0.2$.} 
\end{figure}

Again, we start by computing $\mu$. It is given by
\begin{align}
    \mu = \frac{1}{L} \mathbf E\left[ \sum_{\mu=1}^{K} \sum_{k=1}^{\overline N} \sum_{a=1}^{M} \eta^{\mu a} \overline \eta^{\mu b}_k \overline \xi_k^1 \xi^1 \right].
\end{align}
As in the supervised case, only the term relative to $\mu=1$ survives; $\mu$ then reads:
\begin{align}
    \mu = \frac{1}{L} \sum_{k} \sum_{a} \mathbf E_{\chi} \left[ \chi^{1 a} \right] \mathbf E_{\overline \chi}\left[\overline \chi^{1 a}_k \right] = \frac{1}{L} M \overline N r \bar r = \gamma^{-1} M r \bar r.
\end{align}
We now compute $s$: 
\begin{align}
    \mathbf E \left[\left( h^{(t=0)} \xi^1\right)^2\right] = \frac{1}{L^2} \mathbf E\left[ \left(\sum_{\mu=1}^{K} \sum_{k=1}^{\overline N} \sum_{a=1}^{M} \xi^1 \xi^\mu \bar \xi_k^1 \bar \xi_k^\mu \chi^{\mu a} \overline \chi^{\mu a}_k \right)^2\right],
\end{align}
which can be again decomposed in two contributions, $\mu=1$ and $\mu>1$. The first term ($\mu=1$) then reads:
\begin{align}
    \frac{1}{L^2} \sum_{k,l} \sum_{a,b} \mathbf E\left[ \chi^{1 a} \chi^{1 b} \overline \chi^{1 a}_k \overline \chi^{1 b}_l \right] = \frac{\overline N^2}{L^2} \left(M^2 \bar r^2 r^2 + M(1-r^2) \bar r^2 \right) = \gamma^{-2} M^2 \bar r^2 r^2 \left( 1 + \frac{1-r^2}{Mr^2} \right).
\end{align}
In the second term ($\mu>1$), the only relevant contribution is that of order $\mathcal O (\overline N K)$, which reads:
\begin{align}
    \frac{1}{L^2} \sum_{\mu>1}\sum_{k}\mathbf E \left[ \left( \sum_{a} \xi^\mu \xi^1 \bar \xi^1_k \bar \xi^1_k \chi^{\mu a} \bar \chi^{\mu a}\right)^2\right] = \frac{1}{L^2} \sum_{\mu>1}\sum_{k} \sum_{a,b}\mathbf E_\chi \left[ \chi^{\mu a} \chi^{\mu b} \right] \mathbf E_{\bar \chi}\left[\bar \chi^{\mu a} \bar \chi^{\mu b}\right]. 
\end{align}
It can be easily computed, hence it follows
\begin{align}
    \frac{\overline N K}{L^2} (M^2 r^2 \bar r^2 + M (1-r^2) \bar r^2 + M r^2 (1-\bar r^2) + M (1-r^2)(1-\bar r^2)) = \alpha M^2 \bar r^2 r^2 \left( 1+\frac{1-r^2\bar r^2}{M r^2 \bar r^2} \right).
\end{align}
Collecting all the terms, we arrive at the following expression for $s$:
\begin{align}
    s = \gamma^{-2} M^2 \bar r^2 r^2 \left[ \frac{1-r^2}{Mr^2} + \bar \alpha\left(1+\frac{1-r^2\bar r^2}{M r^2 \bar r^2}\right)\right]
\end{align}
Hence, $m^1_\sigma$ finally reads:
\begin{align}
    m^1_\sigma = \erf\left( \frac{1}{\sqrt{2\left( \frac{1-r^2}{Mr^2} + \bar \alpha\left(1+\frac{1-r^2\bar r^2}{M r^2 \bar r^2}\right)\right)} }\right).
\end{align}
The stability condition for the retrieval by the layer $\sigma$ of the pattern $\xi^1$ is then:
\begin{align}
    \frac{1-r^2}{Mr^2} + \bar \alpha\left(1+\frac{1-r^2\bar r^2}{M r^2 \bar r^2}\right)< 1.
    \label{eq:unsupervised_ThrForLearning_1}
\end{align}
The stability condition for the retrieval by the layer $\tau$ of the pattern $\bar \xi^1$ is
\begin{align}
    \frac{1-\bar r^2}{M\bar r^2} + \alpha\left(1+\frac{1-r^2\bar r^2}{M r^2 \bar r^2}\right)< 1.
    \label{eq:unsupervised_ThrForLearning_2}
\end{align}
\begin{figure}[!ht]    
    \centering
    \includegraphics[width=7.5cm]{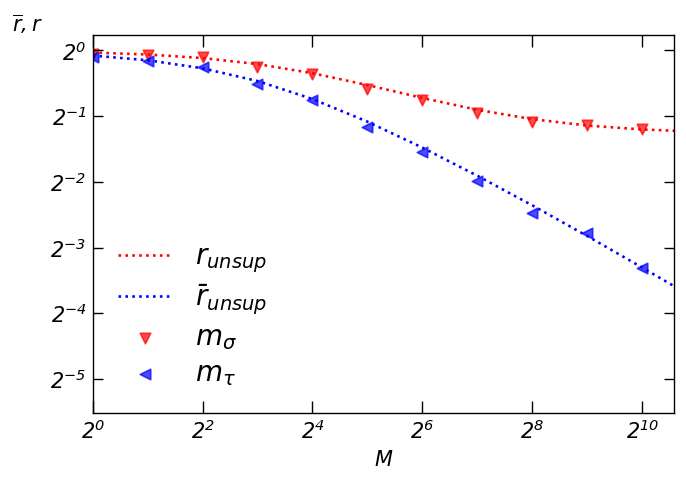}
    \includegraphics[width=7.5cm]{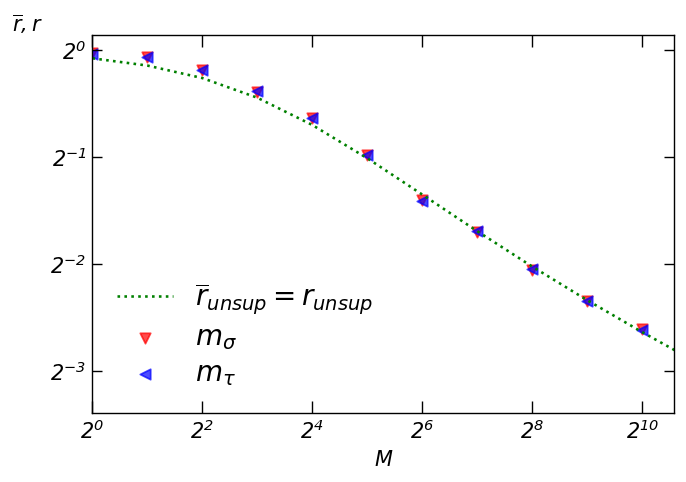}
    \caption{\label{S2N-Unsup} The theoretical predictions derived in Equations \ref{eq:unsupervised_ThrForLearning_1},\ \ref{eq:unsupervised_ThrForLearning_2} are compared to simulated data for two different configurations of the network in the unsupervised scenario: asymmetric (\emph{left} - $N=200, \overline N=100$) and symmetric (\emph{right} - $N=200, \overline N=200$). The number of stored patterns is $K=12$. The plots are obtained by defining a fixed threshold for $m_\sigma$ and $m_\tau$, resulting in a condition on the level of dataset entropy $r$ and $\bar r$ as a function of $M$, as shown in the figures.}
\end{figure}
As shown in Figure \ref{S2N-Unsup}, the above conditions -eq.s (\ref{eq:unsupervised_ThrForLearning_1}) and (\ref{eq:unsupervised_ThrForLearning_2})- provide a recipe to guarantee, before starting to train the networks, if we have enough information to provide to it to ensure the correct inference, storage and successive retrieval of the archetypes over both the layers.  Further, in Figure \ref{Esempi} we report the growth of the magnetizations of the two archetype under retrieval as more and more examples are supplied to the network. 

\bigskip

With a little more effort we can obtain an explicit prescription to evaluate the thresholds for learning, namely the critical scaling of the number of examples to be provided to the network to ensure a correct learning process $M_c(r)$.  We have to distinguish among a symmetric BAM ($\gamma=1$) and the asymmetric case ($\gamma \neq 1$). 
\newline
As far as $\gamma=1$ (that is, we work in the optimal setting where the BAM is equipped with equal size layers and thus the maximal amount of synapses) these result, respectively, in $M_c \sim r^{-2}$ as well as $\bar{M}_c \sim \bar{r}^{-2}$ in the presence of a teacher and they grow up to $M_c \sim r^{-4}$ and $\bar{M}_c \sim \bar{r}^{-4}$ if the teacher is lacking,  as expected \cite{Kanter1,Kanter2,Kanter3,agliari2022emergence}.
\newline
At contrary, if $\gamma \neq 1$, the presence of a teacher is essential as in this scenarios enlarging the dataset size (e.g. providing a larger amount of examples to the network) does not simplify the task so straightforwardly, as it shines by a glance at the left panel of Figure \ref{S2N-Unsup} (red curve): we deepen details of these scalings in the Appendix C.

\section{Conclusions and outlooks}\label{SezioneSei}

In this work we performed a statistical mechanical analysis of the Bidirectional Associative Memories provided with a generalisation of the synaptic matrix that extends hetero-associative Hebbian storage toward hetero-associative Hebbian learning, both in the presence of a teacher (i.e., via a supervised protocol) as well as in its absence (i.e., via an unsupervised protocol).
\newline
There is a number of conclusions that we can state from this analysis:
\begin{itemize}
    \item The integral representation of the partition function related to the BAM network (see Figure \ref{Duality} and eq. (\ref{IntegraZ}))  shows that this machine can be represented by two Restricted Boltzmann machines, whose visible layers are the visible layers of the original BAM, while their hidden layers are made solely by grand-mother cells (that is neurons that selectively fires when a given pattern is presented to the visible layer) in binary communication among the two hidden layers.

    \item Thanks to this duality we can appreciate how retrieval of couples of patterns is achieved, namely we can explain {\em reverberation} by a Machine Learning perspective: as the first RBM is provided with an example -say $\eta^{\nu, a}$- and it is able to recognize the archetype hidden behind the noise (that is $\xi^{\nu}$), its grand-mother cell starts to fire (i.e. $z_{\nu}>>1$), triggering its coupled grand-mother cell in the hidden layer of the other RBM, hence $z_{\nu}^{\dag} >>1$ too, that in turn will succeed in retrieving the correct related archetype $\bar{\xi}^{\nu}$ on the other visible layer. 

    \item The above explanation suggests also that synapses do not share all the same importance as, if we would store couples of patterns in an autoassociative network (i.e. in the Hopfield model) so to account for standard Pavlov's Classical Conditioning, the amount of links to vary would be extensive in the volume, while connecting directly the grand-mother cells allows to save all these synapses (that can thus be used for other tasks). 

    \item We derived explicitly the thresholds for learning, namely we know explicitly and {\em a priori}, given the amount of archetypes that we want the network to handle and the noise at which the network operates, if we have the required information in the datasets to undergo a successful training of the machine or  we do not. Furthermore the critical scalings in the number of examples to provide to the network are power-laws, resulting in $M_c \sim r^{-2}$ in the supervised regime and in  $M_c \sim r^{-4}$ in the unsupervised counterpart (in the symmetric case, asymmetric cases are slightly more subtle). 

    \item The BAM can use these extended resources (provided by two visible layers) to retrieve an archetype on a layer even if provided with a highly entropic dataset by taking advantage of the other layer (if provided with enough information, i.e., if supplied with a low-entropy dataset).

    \item We obtained phase diagrams in the space of the control parameters, that are $\rho, \bar{\rho}$ (or equivalently, $r, M$ and $\bar{r}, \bar{M}$) for the dataset and $\alpha, \beta, \gamma$ for the network's load, noise and architecture: the knowledge of the retrieval region (that is, knowing {\em a priori} for which control parameter values the network will operate as an associative memory), constitutes valuable information for an optimized theory of AI as it prevents useless or conceptually wrong training.
\end{itemize}

Further research extensions of this work would be dedicated to inspect higher order generalizations of these machines, as -for instance- a TAM, namely a three-directional associative memory that stores and retrieves triplets of patterns: we plan to report soon on this topic.

\section*{Acknowledgments}
This work has been supported by Ministero degli Affari Esteri e della Cooperazione Internazionale (MAECI) via the BULBUL grant (Italy-Israel collaboration), CUP Project n. F85F21006230001 {\em  Brain-inspired Ultra-fast and Ultra-sharp machines for assisted healthcare}.
\newline
The PRIN-2022 grant {\em Statistical Mechanics of Learning Machines: from algorithmic and information-theoretical limits to new biologically inspired paradigms} is also acknowledged.

\setcounter{equation}{0}
\renewcommand\theequation{A.\arabic{equation}}

\appendix
\section{Derivation of the Statistical pressure in the supervised scenario with the Guerra's interpolation technique}\label{supAppendix}
In order to keep the calculations as general as possible, we formally define a shorthand for the denominator of the Hamiltonian (\ref{eq:sup}), namely we put $D=M^2 \sqrt{R \overline R}$, and we use a slightly different version of the example magnetizations (\ref{eq:exmagnetizations_supsigma}, \ref{eq:exmagnetizations_suptau}), and of the momentum of the overlaps (\ref{eq:sup_p}, \ref{eq:sup_pbar}), namely
\begin{align}
    &n^{\mu}_\sigma = \frac{1}{NM} \sum_{i=1}^N \lr{\sum_{a=1}^M \eta^{\mu a}_i} \sigma_i,\\
    &n^{\mu}_\tau = \frac{1}{\overline NM} \sum_{k=1}^{\overline N} \lr{\sum_{a=1}^M \bar \eta^{\mu a}_k} \tau_k,\\
    &p^{\alpha\beta} = \frac{1}{K-1} \sum_{\mu>1} z_\mu^\alpha\:z_\mu^\beta,\\
    &\bar p^{\alpha\beta} = \frac{1}{K-1} \sum_{\mu>1} (z^\dag_\mu)^\alpha\:(z^\dag_\mu)^\beta.
\end{align}
To set up the correct interpolation, we introduce the auxiliary real-valued functions $a(t),\ b(t), \bar{b}(t)$ that we use to define a generalized  interpolating partition function that reads as
\begin{align}
    &Z_{N,\bar{N},M}(\beta;t)=\sum_{\{\sigma\} \{\tau\}}\int Dz^\dag Dz \:e^{-\beta W(t)};\\
    &-\beta W(t) = a(t) \frac{\beta L M^2}{D} n^1_\sigma n^1_\tau + b(t) \beta M \sqrt{\frac{R}{ND}} \sum_{i,\mu>1} \lambda^\mu_i z_\mu \sigma_i + \bar b(t) \beta M \sqrt{\frac{\overline R}{\overline ND}} \sum_{k,\mu>1} \bar \lambda^\mu_k z^\dag_\mu \tau_k + L \phi(t),
\end{align}
where we also introduced the function $\phi(t)$, that accounts for a factorized (in both the $\sigma$ and $\tau$ neurons) one-body effective model, whose fields have to suitably mimic the real fields (i.e. the post-synaptic potentials): the explicit expression is provided later on (see  eq. (\ref{OneBodyEffo})) when it will become clear how to write its structure.
\newline
By a glance at eq. (\ref{InterMilan}), we have to evaluate the Cauchy condition $A(\beta; t=0)$ and we must integrate $d_t A(\beta;t)$. Let us perform the computation of the derivative of the free energy: it goes as follows 
\begin{equation}
    \frac{d}{dt} A(\beta; t) = \lim_{L\to\infty} \frac{1}{L} \left[ a'(t) \frac{\beta L M^2}{D} \bavg{n^1_\sigma n^1_\tau}_t + b'(t) \beta M \sqrt{\frac{R}{ND}} \sum_{i,\mu>1} \bavg{\lambda^\mu_i z_\mu \sigma_i}_t + \bar b'(t) \beta M \sqrt{\frac{\overline R}{\overline ND}} \sum_{k,\mu>1} \bavg{\bar \lambda^\mu_k z^\dag_\mu \tau_k}_t + L \bavg{\phi'(t)}_t \right],
    \label{eq:Adot}
\end{equation}
where $a'(t),  b'(t), \bar b'(t), \phi'(t)$ denote the $t-$derivatives of these functions.\\
Writing such a derivative in terms of the order parameters, once performed the asymptotic limit $(N, \bar{N}) \to \infty$, we have

\begin{equation}
    \frac{d}{dt} A(\beta; t) = a' \frac{\beta M^2}{D} \bavg{n^1_\sigma n^1_\tau}_t + b b' \alpha \gamma \frac{ \beta^2 M^2 R}{D} \left( \bavg{p_{11}}_t - \bavg{p_{12} q_{12}}_t \right) + \bar b \bar b' \alpha \gamma \frac{\beta^2 M^2 \overline R}{D} \left( \bavg{\overline p_{11}}_t - \bavg{\overline p_{12} \overline q_{12}}_t \right)+ \bavg{\phi'(t)}_t.
    \label{eq:Adot2}
\end{equation}
The  order parameters $p_{11}, p_{12}, q_{12}, ..$ are naturally introduced  in this derivative as, for instance, shown  by considering  the following expectation (the others work sharply in the same manner):
\begin{align}
     \sum_{i,\mu>1} \bavg{\lambda^\mu_i z_\mu \sigma_i}_t =  \sum_{i,\mu>1} \mathbf{E}_{\lambda} \left[ \lambda^\mu_i \omega_t(z_\mu \sigma_i) \right]=\sum_{i,\mu>1} \mathbf{E}_{\lambda} \left[ \pder[ \omega_t(z_\mu \sigma_i) ]{\lambda^\mu_i}\right],
\end{align}
where the last step has been achieved by virtue of the Wick's theorem: $\int Dx\: x f(x,y) = \int Dx \pder[f(x,y)]{x}$, with $Dx$ denoting the Gaussian measure of x. Thus, after computing the derivative in the last passage we can write
\begin{align}
     \sum_{i,\mu>1} \mathbf{E}_{\lambda} \left[ \pder[ \omega_t(z_\mu \sigma_i) ]{\lambda^\mu_i}\right] = \frac{b(t) \beta M R^{1/2}}{\sqrt{ND}} \:\sum_{i,\mu>1} \mathbf{E}_{\lambda} \left[ \omega_t((z_\mu \sigma_i)^2) - \omega^2_t(z_\mu \sigma_i) \right];
\end{align}
notice that $\sigma_i^2 = 1$. The last step is to write the two averages $\omega_t(z_\mu^2)$ and $\omega_t^2(z_\mu \sigma_i)$ as:
\begin{align}
    &\sum_{i,\mu>1}\omega_t(z_\mu^2) = N\omega_t\left(\sum_{\mu>1} z_\mu^{(1)} z_\mu^{(1)}\right) = NK \omega_t(p_{11})\\
    &\sum_{i,\mu>1}\omega^2_t(z_\mu \sigma_i) = \omega_t\left(\sum_{\mu>1} z_\mu^{(1)} z_\mu^{(2)} \sum_{i}\sigma^{(1)}_i \sigma^{(2)}_i\right) = K N \omega_t(p_{12} q_{12}),
\end{align}
where we used the definitions of $p^{ab}$ and $q^{ab}$ and we wrote $K$ instead of $K-1$ for simplicity (as $K\to \infty$ in the present theory). The computation is much the same  for all the other term in Eq. (\ref{eq:Adot}), leading to Eq. (\ref{eq:Adot2}).\\
The role of the function $\phi(t)$ is to balance the terms arising from the derivative of $A$ which are difficult to compute explicitly. In the thermodynamic limit $L\to \infty$ the following quantities vanish under the assumption of replica symmetry:
\begin{align}
    &\bavg{\Delta n^1_\sigma \Delta n^1_\tau} \equiv \bavg{(n^1_\sigma - n_\sigma) (n^1_\tau - n_\tau)}\\
    &\bavg{\Delta p \Delta q} \equiv \bavg{(p^{12} - p) (q^{12} - q)}\\
    &\bavg{\Delta \bar p \Delta \bar q} \equiv \bavg{(\bar p^{12} - \bar p) (\bar q^{12} - \bar q)}   
\end{align}
where with $n_\sigma, n_\tau$ we indicate the expected value of the example magnetizations $n^1_\sigma, \: n^1_\tau$, \emph{i.e.} $\bavg{n^1_\sigma}=n_\sigma, \: \bavg{n^1_\tau}=n_\tau$, similarly for the overlaps $\bavg{q^{12}}=q,\: \bavg{\bar q^{12}}=\bar q, \: \bavg{p^{12}}=p, \:\bavg{\bar p^{12}}=\bar p$. Hence, the idea is to write down $\phi$ as a linear combination of quadratic and one-body terms which are easy to integrate. It reads:
\begin{align}\label{OneBodyEffo}
    L \phi(t) = c(t) \sum_{\mu>1} z_\mu^2 + \bar c(t) \sum_{\mu>1} (z^\dag_\mu)^2 + u(t) \sum_{\mu>1} \rho_\mu z_\mu + \bar u(t) \sum_{\mu>1} \bar \rho_\mu z^\dag_\mu &+ v(t) \sum_{i} h_i \sigma_i + \bar v(t) \sum_{k} \bar h_k \tau_k +\nonumber\\
    &+w(t) \sum_{i,a} \xi^1_i \chi^{1a}_i \sigma_i + \bar w(t) \sum_{k,a} \bar \xi^1_k \bar \chi^{1a}_k \tau_k ,
\end{align}
where we introduced the new unknown real-valued functions $c(t), \bar c(t), u(t), \bar u(t), v(t), \bar v(t), w(t), \bar w(t)$. The combined average of the $t-$derivative of $\phi$ in Eq. (\ref{eq:Adot2}) reads
\begin{align}
    \bavg{\phi'(t)}_t = c' \alpha \gamma \bavg{p_{11}}_t + \bar c' \alpha \gamma \bavg{p_{11}}_t &+ u u' \alpha \gamma \left( \bavg{p_{11}}_t - \bavg{p_{12}}_t \right) + \bar u \bar u' \alpha \gamma \left( \bavg{\bar p_{11}}_t - \bavg{\bar p_{12}}_t \right)+\nonumber\\
    & + v v' \gamma \left(1- \bavg{q_{12}}_t \right) + \bar v \bar v' \gamma^{-1} \left(1- \bavg{\bar q_{12}}_t \right) + w' \gamma M \bavg{n^1_\sigma}_t + \bar w' \gamma^{-1} M \bavg{n^1_\tau}_t.
    \label{eq:phidot}
\end{align}
Combining Eqs. (\ref{eq:phidot}) with Eq. (\ref{eq:Adot2}), the $t-$derivative of $A$ reads:
\begin{align}
    \frac{d}{dt} A(\beta; t) = &a' \frac{\beta M^2}{D} \bavg{n^1_\sigma n^1_\tau}_t + b b' \alpha \gamma \frac{ \beta^2 M^2 R}{D} \left( \bavg{p_{11}}_t - \bavg{p_{12} q_{12}}_t \right) + \bar b \bar b' \alpha \gamma \frac{\beta^2 M^2 \overline R}{D} \left( \bavg{\overline p_{11}}_t - \bavg{\overline p_{12} \overline q_{12}}_t \right)+\nonumber\\
    &+ c' \alpha \gamma \bavg{p_{11}}_t + \bar c' \alpha \gamma \bavg{p_{11}}_t + u u' \alpha \gamma \left( \bavg{p_{11}}_t - \bavg{p_{12}}_t \right) + \bar u \bar u' \alpha \gamma \left( \bavg{\bar p_{11}}_t - \bavg{\bar p_{12}}_t \right)+\nonumber\\
    & + v v' \gamma \left(1- \bavg{q_{12}}_t \right) + \bar v \bar v' \gamma^{-1} \left(1- \bavg{\bar q_{12}}_t \right) + w' \gamma M \bavg{n^1_\sigma}_t + \bar w' \gamma^{-1} M \bavg{n^1_\tau}_t.
\end{align}
Collecting the homogeneous terms in the above equation and systematically eliminating the terms containing the expectation values of the order parameters, \emph{i.e.} $\bavg{q^{12}}_t, \bavg{\bar q^{12}}_t, \bavg{p^{12}}_t, \bavg{\bar p^{12}}_t, \bavg{n^1_\sigma}_t, \bavg{n^1_\tau}_t$, one defines a system of differential equations for the functions $a,b,\bar b, c, \bar c, u, \bar u, v, \bar v, w, \bar w$:
\begin{align}
    &c'+u u'+ \frac{\beta^2 M^2 R}{D} b b' = 0,\label{eq:pdc0}\\
    &\bar c'+\bar u \bar u'+ \frac{\beta^2 M^2 \overline R}{D} \bar b \bar b' = 0,\\
    & u u' + \frac{\beta^2 M^2 R q}{D}  b b' = 0,\\
    &\bar u \bar u' + \frac{\beta^2 M^2 \overline R \bar q}{D}  \bar b \bar b' = 0,\\
    &v v' + \frac{\beta^2 \alpha M^2 R p}{D} b b' = 0,\\
    &\bar v \bar v' + \frac{\beta^2 \alpha \gamma^2 M^2 \overline R \bar p}{D} b b' = 0,\\
    &w' + a'\frac{\beta M}{ D} n_\tau \gamma^{-1} = 0,\\
    &\bar w' + a'\frac{\beta M}{D} n_\sigma \gamma= 0,
    \label{eq:pdc}
\end{align}
with boundary conditions
\begin{align}
    &a(t=0) = b(t=0) = \bar b(t=0)=0,\\
    &a(t=1) = b(t=1) = \bar b(t=1)=1,\\
    &\phi(t=1)=0.
\end{align}
Without loss of generality, one can impose the following constraints on the system (\ref{eq:pdc0})-(\ref{eq:pdc}):
\begin{align}
    &a' = 1,\\
    &b b' = \bar b \bar b'  = 1/2,
\end{align}
such that the overall solution reads 
\begin{align}
    &a(t)=t,\\
    &b(t)=\bar b (t) = \sqrt{t},\\
    &c(t)=\frac{1}{2} \frac{\beta^2 M^2 R}{D} (1-q)(1-t),\\
    &\bar c(t)=\frac{1}{2} \frac{\beta^2 M^2 \overline R}{D} (1-\bar q)(1-t),\\
    &u(t)= \sqrt{\frac{\beta^2 M^2 R}{D}q(1-t)},\\
    &\bar u(t)= \sqrt{\frac{\beta^2 M^2 \overline R}{D}\bar q(1-t)},\\
    &v(t)= \sqrt{\frac{\beta^2 \alpha M^2 R}{D}p(1-t)},\\
    &\bar v(t)= \sqrt{\frac{\beta^2 \alpha \gamma^2 M^2 \overline R}{D}\bar p (1-t)},\\
    &w(t) = n_\tau \gamma^{-1} \frac{\beta M}{D} (1-t),\\
    &\bar w(t) = n_\sigma \gamma \frac{\beta M}{D} (1-t).    
\end{align}
Finally, by imposing the  replica symmetric \emph{RS} ansatz (that is neglecting fluctuations of the order parameters around their means), the $t-$derivative of the interpolating quenched pressure $A(\beta;t)$ becomes
\begin{align}
    \frac{d}{dt}A_{RS}(\beta;t) = -\beta \frac{M^2}{D} n_\sigma n_\tau - \frac{1}{2} \frac{\beta^2 M^2}{D} \alpha \gamma \left( R p (1-q) + \overline R \bar p (1-\bar q)\right).
\end{align}
Note that such a derivative does not depend any longer on $t$, thus the quenched pressure $A_{RS}(\beta)$ reads
\begin{equation}
    A_{RS}(\beta)= A_{RS}(\beta; t=1) = A_{RS}(\beta; t=0) + \frac{d}{dt} A_{RS}(\beta; t),
\end{equation}
as integrating a constant in $t$ for $t \in (0,1)$ is nothing but multiplication by one. 
\newline
We can finally evaluate the Cauchy condition $A_{RS}(\beta; t=0)$. We first write explicitly $Z(\beta;t=0)$, which reads
\begin{align}
    Z(\beta;t=0) = \sum_{\{\sigma\} \{\tau\}}\int Dz^\dag Dz \:\exp\left(-\beta W(t=0)\right)
\end{align}
with (indicating the $a, b, \bar b, ..$ function evaluated at $t=0$ as $a_0, b_0, \bar b_0 ,..$):
\begin{align}
    -\beta W(t=0) = c_0 \sum_{\mu>1} z_\mu^2 + \bar c_0 \sum_{\mu>1} (z^\dag_\mu)^2 + u_0 \sum_{\mu>1} \rho_\mu z_\mu + \bar u_0 \sum_{\mu>1} \bar \rho_\mu z^\dag_\mu &+ v_0 \sum_{i} h_i \sigma_i + \bar v_0 \sum_{k} \bar h_k \tau_k +\nonumber\\
    &+w_0 \sum_{i,a} \xi^1_i \chi^{1a}_i \sigma_i + \bar w_0 \sum_{k,a} \bar \xi^1_k \bar \chi^{1a}_k \tau_k .
\end{align}
The partition function $Z(\beta;t=0)$ reads
\begin{align}
    Z(\beta;t=0) = \left[2\cosh\left( v_0 h + w_0 \sum_a \xi^1 \chi^{1a} \right)\right]^{N} &\left[2 \cosh \left( \bar v_0 \bar h + \bar w_0 \sum_a \bar \xi^1 \bar \chi^{1a} \right) \right]^{\overline N}\times\nonumber\\
    &\times \prod_{\mu>1} \int d^2 X_{\mu} \exp\left( -\frac{1}{2} X^T_{\mu} C X_{\mu} + X^T_{\mu} Y_{\mu} \right)
    \label{eq:Z0}
\end{align}
where we introduced the vectors $X_{\mu}, Y_{\mu}$ and the matrix $C$ as follows:
\begin{align}
    X_{\mu}=&
    \begin{bmatrix}
        \mathcal Re (z_{\mu})\\
        \mathcal Im (z_{\mu})
    \end{bmatrix},\:\:\:
    Y_{\mu}=
    \begin{bmatrix}
        u_0 t_{\mu} + \bar u_0 \bar t_{\mu} \\
        i(u_0 t_{\mu} - \bar u_0 \bar t_{\mu})
    \end{bmatrix}, \:\:\:
    C =
    \begin{bmatrix}
        \beta - (c_0+\bar c_0) & i (\bar c_0 - c_0)\\
        i (\bar c_0 - c_0) & \beta + (c_0+\bar c_0)
    \end{bmatrix}.
\end{align}
The Gaussian integral in Eq. (\ref{eq:Z0}) can be easily computed: it reads
\begin{align}
    \frac{(2\pi)^K}{(\det C)^{K/2}} \prod_{\mu>1} \exp\left( \frac{1}{2} Y^T_{\mu} C^{-1} Y_{\mu} \right),
    \label{eq:exponent}
\end{align}
where the determinant of $C$ is
\begin{align}
    \det C = 4(\beta^2 - 4 c_0 \bar c_0 )= 4\beta^2\left[ 1-\beta^2 \frac{R \overline R M^4}{D^2}(1-q)(1-\bar q)\right].
\end{align}
The quenched expectation of the exponent in (\ref{eq:exponent}) reads
\begin{align}
    \avg{Y^T_{\mu} C^{-1} Y_{\mu}}_{\rho, \bar \rho}=\frac{4 \beta^4}{\det C}  \frac{R \overline R M^4}{D^2} \left( q(1-\bar q) + \bar q (1-q)\right),
\end{align}
such that, finally, we end up with the following expression for $A_{RS}$:
\begin{align}
    A_{RS}= &-\beta \frac{M^2}{D} n_\sigma n_\tau - \frac{1}{2} \frac{\beta^2 M^2}{D} \alpha \gamma \left( R p (1-q) + \overline R \bar p (1-\bar q)\right)+\nonumber\\
    &-\frac{\alpha \gamma}{2} \ln \left( 1-\beta^2 \frac{R \overline R M^4}{D^2}(1-q)(1-\bar q) \right) + \frac{\alpha \gamma \beta^2}{2} \frac{R \overline R M^4}{D^2} \frac{ q(1-\bar q) + \bar q (1-q)}{1-\beta^2 \frac{R \overline R M^4}{D^2}(1-q)(1-\bar q)}+\nonumber\\
    &+\gamma \mathbf{E}_{x,\xi, \chi} \ln \cosh \left( \beta \sqrt{\alpha \frac{R M^2}{D} p} \:x + \beta n_\tau \gamma^{-1} \left( \frac{M}{D} \sum_a \chi_a\right) \xi \right)+\nonumber\\
    &+\gamma^{-1} \mathbf{E}_{x,\bar \xi, \bar \chi} \ln \cosh \left( \beta \sqrt{\bar \alpha \frac{\overline R M^2}{D} \bar p}\: x + \beta n_\sigma \gamma \left( \frac{M}{D} \sum_a \bar \chi_a\right) \bar \xi \right),
    \label{eq:ARS1}
\end{align}
where $\mathbf{E}_{x,\xi, \chi}= \mathbf{E}_{\xi, \chi}\int Dx$, and $\mathbf{E}_{x,\bar \xi,\bar \chi}= \mathbf{E}_{\bar \xi,\bar \chi}\int Dx$, are the expectations operated on the Gaussian variable $x\sim \mathcal N (0,1)$ and on the archetypes $\xi, \bar \xi$ and on $ \chi, \bar \chi$.\\
Reintroducing the definition of $D$ in the latter equation, $D=M^2 \sqrt{R\overline R}$ and using the rescaled definition of the order parameters $n_\sigma, n_\tau, p, \bar p$, and after computing the expectations on $\xi$ and $\bar \xi$ using the symmetry of the Gaussian integrals on $x$, we finally derive eq. \ref{eq:ARS2}.

\setcounter{equation}{0}
\renewcommand\theequation{B.\arabic{equation}}

\section{Appendix: derivation of the Statistical pressure in the unsupervised scenario with the Guerra's interpolation technique}\label{unsupAppendix}

We show here the computation of the \emph{RS}-statistical pressure $A_{RS}$ (eq. \ref{eq:Adefinition}) using the Guerra's interpolation technique (eq.\ref{InterMilan}). The details of the method have been reported in the previous section, hence here we highlight the main results.\\
We start by introducing a short-hand definition of the denominator of the unsupervised hamiltonian, namely $D=\sqrt{R \overline R}$ and, as in the supervised case, we work with the simplified version of the example magnetizations $n^a_\sigma,n^a_\tau$, which will be rescaled at the end in order to match the definition given in equations \ref{eq:exmagnetizations_unsupsigma},\ref{eq:exmagnetizations_unsuptau}. The simplified example magnetizations are
\begin{align}
    &n^{\mu a}_\sigma = \frac{1}{N} \sum_{i=1}^N  \eta^{\mu a}_i \sigma_i,\\
    &n^{\mu a}_\tau = \frac{1}{\overline N} \sum_{k=1}^{\overline N} \bar \eta^{\mu a}_k \tau_k.
\end{align}
In these terms, the parametric partition function reads:
\begin{align}
    Z(t)=\sum_{\sigma,\tau}\exp\lr{ a(t)\frac{\beta L}{DM} \sum_a n^{1a}_\sigma n^{1a}_\tau + b(t)r \bar r \sqrt{1+\rho} \frac{\beta}{LDM} \sum_{\mu>1} \sum_{ik} z^\mu_{ik}\sigma_i \tau_k + L\phi(t)}.
\end{align}
We start the calculation by evaluating the $t-$derivative of $A$ is
\begin{align}
    \der[A(\beta;t)]{t}=\lim_{L\to \infty}\frac{1}{L}\lr{a' \frac{\beta L}{MD} \sum_a \bavg{n^{1a}_\sigma n^{1a}_\tau} + b b' r^2 \bar r^2 (1+\rho) K \frac{\beta^2}{D^2} (1-\bavg{q_{12}\bar q_{12}} + L\bavg{\phi'}}
\end{align}
where we have introduce $\phi$ as 
\begin{align}
    L\phi(t)=v(t) \sum_i h_i \sigma_i + \bar v(t) \sum_k \bar h_k \tau_k + w(t) \sum_{ia} \eta^{1a}_{a} \sigma_i + \bar w(t) \sum_{ka} \bar \eta^{1a}_{a} \tau_k.
\end{align}
The $t-$derivative of $\phi$ is easily computed and reads
\begin{align}
    L\bavg{\phi'}=v v' N (1-\bavg{q_{12}})+\bar v \bar v' \bar N (1-\bavg{\bar q_{12}}) + w' N \sum_a \bavg{n^{1a}_\sigma} + \bar w' \overline N \sum_a \bavg{n^{1a}_\tau}.
\end{align}
Hence, the $t-$derivative of the quenched statistical pressure is given by
\begin{align}\label{eq:unsupAdot}
    \der[A(\beta;t)]{t}=&a' \frac{\beta}{MD} \sum_a \bavg{n^{1a}_\sigma n^{1a}_\tau} + w' \gamma \sum_a \bavg{n^{1a}_\sigma} + \bar w' \gamma^{-1} \sum_a \bavg{n^{1a}_\tau} +\nonumber\\
    +&b b' r^2 \bar r^2 (1+\rho) \lambda \frac{\beta^2}{D^2} (1-\bavg{q_{12}\bar q_{12}}) + v v' \gamma (1-\bavg{q_{12}})+\bar v \bar v' \gamma^{-1} (1-\bavg{\bar q_{12}}). 
\end{align}
We have divided the latter equation in two lines; the first line contain the signal terms that have to be balanced by an appropriate choice of the functions $a(t),w(t),\bar w(t)$; the second line is relative to the background noise exerted by the condensed patterns $\eta^{\mu,a}$ for $\mu>1$, which is treated separately.\\
We begin by analysing the first line. Under the following definition of the functions $a,w,\bar w$:
\begin{align}
    &a(t)=t,\\
    &w(t)=\beta\frac{n_\tau \gamma^{-1}}{DM}(1-t),\\
    &\bar w(t)=\beta\frac{n_\sigma \gamma}{DM}(1-t),\\
\end{align}
the first line in eq. \ref{eq:unsupAdot} can be written as
\begin{align}
    &\frac{a' \beta}{MD} \sum_a \lr{\bavg{n^{1a}_\sigma n^{1a}_\tau} + \frac{w' \gamma}{a' \beta} DM \bavg{n^{1a}_\sigma} + \frac{\bar w' \gamma^{-1} }{a' \beta} DM \bavg{n^{1a}_\tau}}=\frac{\beta}{D} \bavg{\Delta n^{1a}_\sigma \Delta n^{1a}_\tau} - \frac{\beta}{D} n_\sigma n_\tau
\end{align}
where we have introduced the fluctuations $\Delta n^{1a}_\sigma,\Delta n^{1a}_\tau$ and average example magnetizations $n_\sigma,n_\tau$:
\begin{align}
    &\Delta n^{1a}_\sigma= n^{1a}_\sigma- n_\sigma,\\
    &\Delta n^{1a}_\tau= n^{1a}_\tau- n_\tau,\\
    &n_\sigma = \bavg{n^{1a}_\sigma},\\
    &n_\tau =  \bavg{n^{1a}_\tau}.
\end{align}
We proceed similarly for the second line of eq. \ref{eq:unsupAdot}. First, we introduce the fluctuations and averaged overlaps $\bar q,q$:
\begin{align}
    &\Delta q= q_{12}- q,\\
    &\Delta \bar q= \bar q_{12} - \bar q,\\
    &q = \bavg{q_{12}},\\
    &\bar q =  \bavg{\bar q_{12}}.
\end{align}
Then, we define the functions $b,v,\bar v$ as:
\begin{align}
    &b(t) = \sqrt{t},\\
    &v(t)=\frac{\beta r \bar r}{D}\sqrt{(1+\rho) \lambda \gamma^{-1} \bar q(1-t)},\\
    &\bar v(t)=\frac{\beta r \bar r}{D}\sqrt{(1+\rho) \lambda \gamma q(1-t)}.\\
\end{align}
Under these hypotheses, the second line of eq. \ref{eq:unsupAdot} reads
\begin{align}
    &b b' r^2 \bar r^2 (1+\rho) \lambda \frac{\beta^2}{D^2} (1-\bavg{q_{12}\bar q_{12}}) + v v' \gamma (1-\bavg{q_{12}})+\bar v \bar v' \gamma^{-1} (1-\bavg{\bar q_{12}}) = \frac{1}{2} \frac{\beta^2}{D^2} r^2 \bar r^2 (1+\rho) \lambda \left[(1-q)(1-\bar q)- \bavg{\Delta q \Delta \bar q}\right].
\end{align}
We are now able to compute $A(\beta; t=0)$. The partition function $Z(\beta;t=0)$ reads
\begin{align}
    Z(\beta;t=0)=&\sum_{\sigma,\tau} \exp\lr{v_0 \sum_i h_i \sigma_i + \bar v_0 \sum_k \bar h_k \tau_k + w_0 \sum_{ia} \eta^{1a}_i \sigma_i + \bar w_0 \sum_{ka} \bar \eta^{1a}_k \tau_k } = \\
    =&\:2^{N+\bar N} \cosh \lr{v_0 h + w_0 \sum_a \eta^{1a}} \cosh \lr{\bar v_0 \bar h + \bar w_0 \sum_a \bar \eta^{1a}},
\end{align}
with
\begin{align}
    &v_0 = v(t=0) = \frac{\beta r \bar r}{D}\sqrt{(1+\rho) \lambda \gamma^{-1} \bar q}\\
    &\bar v_0 = \bar v(t=0) = \frac{\beta r \bar r}{D}\sqrt{(1+\rho) \lambda \gamma q}\\
    &w_0 = w(t=0) = \beta\frac{n_\tau \gamma^{-1}}{DM}\\
    &\bar w_0 = \bar w(t=0) = \beta\frac{n_\sigma \gamma}{DM}.    
\end{align}
Hence:
\begin{align}
    A(\beta;t=0)=\gamma \mathbf E \ln \cosh \lr{v_0 h + w_0 \sum_a \eta^{1a}} + \gamma^{-1} \mathbf E \ln \cosh \lr{\bar v_0 \bar h + \bar w_0 \sum_a \bar \eta^{1a}}.
\end{align}
In the RS approximation the fluctuations of the order parameters vanish by definition. We are now able to write the statistical pressure as
\begin{align}
    A_{RS} =& - \frac{\beta}{D} n_\sigma n_\tau + \frac{1}{2} \frac{\beta^2}{D^2} r^2 \bar r^2 (1+\rho) \lambda (1-q)(1-\bar q)+\\
    +&\gamma \mathbf E_{x,\eta} \ln \cosh \lr{\frac{\beta r \bar r}{D}\sqrt{(1+\rho) \lambda \gamma^{-1} \bar q}\: x + \beta\frac{n_\tau \gamma^{-1}}{DM} \sum_a \eta^{1a}} + \\
    +&\gamma^{-1} \mathbf E_{x,\bar \eta} \ln \cosh \lr{\frac{\beta r \bar r}{D}\sqrt{(1+\rho) \lambda \gamma q} \: \bar x + \beta\frac{n_\sigma \gamma}{DM} \sum_a \bar \eta^{1a}}.
\end{align}
After rescaling the example magnetizations, following their definition in equations (\ref{eq:exmagnetizations_unsupsigma},\ref{eq:exmagnetizations_unsuptau}), and rewriting $D=\sqrt{R\overline R}$, we obtain eq. (\ref{eq:ARSunsup}).

\setcounter{equation}{0}
\renewcommand\theequation{C.\arabic{equation}}

\section{Appendix: the threshold for learning}

In this appendix we provide the detailed calculations that lead to the critical thresholds for learning in the two protocols.

\subsection*{Supervised setting}
The stability condition for the retrieval by the layer $\sigma$ of the pattern $\xi^1$ is:
\begin{align} 
    \frac{1-r^2}{Mr^2} + \bar \alpha \left(1+\frac{1-r^2}{Mr^2}\right)\left(1+\frac{1-\bar r^2}{M\bar r^2}\right)< x_1 = \frac{1}{2 \left(\erf^{-1}(\theta_1)\right)^2}.
    \label{eq:supervised_ThrForLearning_1bis}
\end{align}
The stability condition for the retrieval by the layer $\tau$ of the pattern $\bar \xi^1$ is
\begin{align} 
    \frac{1-\bar r^2}{M\bar r^2} + \alpha \left(1+\frac{1-r^2}{Mr^2}\right)\left(1+\frac{1-\bar r^2}{M\bar r^2}\right)< x_2 = \frac{1}{2 \left(\erf^{-1}(\theta_2)\right)^2}.
    \label{eq:supervised_ThrForLearning_2bis}
\end{align}
The critical values of $r_c$ and $\bar r_c$ are given by:
\begin{align}
    &r_c = \frac{1}{\sqrt{1+\rho_0 M}},\\
    &\bar r_c = \frac{1}{\sqrt{1+\bar \rho_0 M}},
\end{align}
where
\begin{align}
    &\rho_0 = -\frac{A}{2} + \sqrt{\frac{A^2}{4} -b},\label{eq:rho}\\
    &\bar\rho_0 = -\frac{\bar A}{2} + \sqrt{\frac{\bar A^2}{4} -\bar b},\label{eq:rhobar}
\end{align}
with:
\begin{align}
    &A=a + c, \:\: a=\gamma^2 x_2 - x_1, \:\: c = \alpha^{-1}+\gamma^2 + 1;\label{eq:ac}\\
    &\bar A=\bar a + \bar c, \:\: \bar a=\gamma^{-2} x_2 - x_1, \:\: \bar c = \bar \alpha^{-1}+\gamma^{-2} + 1;\label{eq:barac}\\
    &b=\gamma^2 x_2 - x_1 + \gamma^2 - \alpha^{-1} x_1 = a + \gamma^2 - \alpha^{-1} x_1;\label{eq:b}\\
    &\bar b=\gamma^{-2} x_2 - x_1 + \gamma^{-2} - \bar \alpha^{-1} x_1 = \bar a + \gamma^{-2} - \bar \alpha^{-1} x_1.\label{eq:barb}
\end{align}
Since the parameters $A,\bar A, b, \bar b$ are intensive in M, the scaling of $r_c$ and $\bar r_c$ is:
\begin{align}
    r_c \sim M^{-1/2};\\
    \bar r_c \sim M^{-1/2}.
\end{align}

\subsection*{Unupervised setting}
The stability condition for the retrieval by the layer $\sigma$ of the pattern $\xi^1$:
\begin{align}
    \frac{1-r^2}{Mr^2} + \bar \alpha\left(1+\frac{1-r^2\bar r^2}{M r^2 \bar r^2}\right)< x_1 = \frac{1}{2 \left(\erf^{-1}(\theta_1)\right)^2}.
    \label{eq:unsupervised_ThrForLearning_1}
\end{align}
The stability condition for the retrieval by the layer $\tau$ of the pattern $\bar \xi^1$ is
\begin{align}
    \frac{1-\bar r^2}{M\bar r^2} + \alpha\left(1+\frac{1-r^2\bar r^2}{M r^2 \bar r^2}\right)< x_2 = \frac{1}{2 \left(\erf^{-1}(\theta_2)\right)^2}.
    \label{eq:unsupervised_ThrForLearning_2}
\end{align}
The critical values of $r_c$ and $\bar r_c$ are again given by:
\begin{align}
    &r_c = \frac{1}{\sqrt{1+\rho_0 M}},\\
    &\bar r_c = \frac{1}{\sqrt{1+\bar \rho_0 M}},
\end{align}
where
\begin{align}
    &\rho_0 = -\frac{A}{2} + \sqrt{\frac{A^2}{4} -B},\label{eq:rho0unsup}\\
    &\bar\rho_0 = -\frac{\bar A}{2} + \sqrt{\frac{\bar A^2}{4} -\bar B}\label{eq:barrho0unsup},
\end{align}
with parameters
\begin{align}
    &A=a + \frac{c}{M},\\
    &\bar A=\bar a + \frac{\bar c}{M},\\
    &B=\frac{b}{M},\\
    &\bar B=\frac{\bar b}{M}.
\end{align}
The parameters $a,\bar a, b, \bar b, c, \bar c$ are given by the definitions in equations \ref{eq:ac},\ref{eq:b},\ref{eq:barac},\ref{eq:barb}. The difference with the supervised case is that now the parameters are no longer intensive in $M$. We now derive asymptotic expansions for $r_c$ and $\bar r_c$. We first notice that the quantity $\frac{4B}{A^2}$ has two different possible scalings:
\begin{align}
    &\frac{4B}{A^2} = \frac{4Mb}{(aM+c)^2}\sim M, \:\:a = 0\\
    &\frac{4B}{A^2} = \frac{4Mb}{(aM+c)^2}\sim M^{-1}, \:\:a \neq 0.
\end{align}
In the natural case where $x_1=x_2$ (we use the same threshold for both layers), the condition on $a$ becomes: $a=0 \Rightarrow \gamma = 1$ (conversely, $a\neq 0 \Rightarrow \gamma \neq 1$). Hence, in the case $\gamma = 1$ ($a=0$) we have:
\begin{align}
    M \rho_0 = -\frac{c}{2} + \frac{c}{2}\sqrt{1-\frac{4b}{c^2}M} \sim M^{1/2}, 
\end{align}
and the same for $\bar \rho_0$, hence the scaling of $r_c$ and $\bar r_c$ reads:
\begin{align}
    r_c \sim M^{-1/4},\\
    \bar r_c \sim M^{-1/4}.
\end{align}
In the case of $\gamma\neq 1$ we can expand square root in eq. \ref{eq:rho0unsup} and \ref{eq:barrho0unsup} w.r.t the parameters $4B/A$ and $4\bar B/\bar A$ respectively. We start with $\rho_0$, in the hypothesis $\gamma>1$, that ensures that $a>0$. We have:
\begin{align}
    \rho_0 \sim -\frac{A}{2} + \frac{A}{2}\left(1-\frac{2B}{A^2}\right) = -\frac{B}{A}=-\frac{b}{c+aM}, \:\:b<0.
\end{align}
Hence we have
\begin{align}
    r_c \sim \frac{1}{\sqrt{1-\frac{bM}{c+aM}}}\sim M^0, \:\: M\to \infty.
\end{align}
In order to obtain the asymptotic behaviour of $\bar r_c$ we have to be careful about the fact that, for $\gamma>1$, $\bar a<0$: for this reason, in order to have $\bar \rho_0>0$ we have to take the other root of the equation \ref{eq:unsupervised_ThrForLearning_2}, namely:
\begin{align}
    &\bar\rho_0 = -\frac{\bar A}{2} - \sqrt{\frac{\bar A^2}{4} -B}\label{eq:barrho0unsup_m}.
\end{align}
We can now expand as before, and we have:
\begin{align}
    \bar \rho_0 \sim -\bar A + \frac{\bar B}{\bar A}=\frac{\bar b M - (c+\bar a M)^2}{(c+\bar a M) M}\sim - \bar a M,\:\:M\gg 1.
\end{align}
Thus, the asymptotic behaviour of $\bar r_c$ is:
\begin{align}
    \bar r_c \sim \frac{1}{\sqrt{1-\bar a M}}\sim M^{-1/2}.
\end{align}
The opposite case $\gamma<1$ naturally follows. In summary we are able to distinguish among three cases, depending on the value of $\gamma$, namely:
\begin{enumerate}
    \item $\gamma<1$: $r_c\sim M^{-1/2}$, $\bar r_c\sim M^0$.
    \item $\gamma=1$: $r_c\sim M^{-1/4}$, $\bar r_c\sim M^{-1/4}$;
    \item $\gamma>1$: $r_c\sim M^0$, $\bar r_c\sim M^{-1/2}$;
\end{enumerate}

\printbibliography

\end{document}